\documentclass[apj]{emulateapj}

\def \agile {AGILE}
\def \egret {EGRET}
\def \glast {{\it Fermi}}

\def \swi {{\it Swift}}

\def \webt {WEBT}

\def \rxte {RXTE}

\def \ergcmsec{\hbox{erg cm$^{-2}$ s$^{-1}$}}
\def \phcmsec{\hbox{photons cm$^{-2}$ s$^{-1}$}}

\def \gray {$\gamma$-ray }
\def \source {\hbox{3C~454.3}}

\shorttitle{Eighteen months of \agile{} monitoring of 3C~454.3}
\shortauthors{Vercellone et al.}

\begin{document}
\title{Multiwavelength observations of 3C~454.3. \\
  III. Eighteen months of \agile{} monitoring of the ``{\it Crazy Diamond}''}

\author{S.~Vercellone\altaffilmark{1,*},
  F.~D'Ammando\altaffilmark{2,3},
  V.~Vittorini\altaffilmark{2,4},
  I.~Donnarumma\altaffilmark{2},
  G.~Pucella\altaffilmark{5},
  M.~Tavani\altaffilmark{2,3,4},
  A.~Ferrari\altaffilmark{4,6},
%
% Multi-lambda Team
%
  C.M.~Raiteri\altaffilmark{7},
  M.~Villata\altaffilmark{7},
  P.~Romano\altaffilmark{1},
  H.~Krimm\altaffilmark{8,9}
% ...
}
\altaffiltext{1}{INAF/IASF--Palermo, Via U.~La Malfa 153, I-90146 Palermo, Italy}
\altaffiltext{2}{INAF/IASF--Roma, Via del Fosso del Cavaliere 100, 
  I-00133 Roma, Italy}
\altaffiltext{3}{Dip. di Fisica, Univ. ``Tor Vergata'', Via della Ricerca 
  Scientifica 1, I-00133 Roma, Italy}
\altaffiltext{4}{CIFS--Torino, Viale Settimio Severo 3, I-10133, Torino, Italy}
\altaffiltext{5}{ENEA--Roma, Via E. Fermi 45, I-00044 Frascati (Roma), Italy}
\altaffiltext{6}{Dip. di Fisica , Univ. di Torino, Via P. Giuria 1,
  I-10125 Torino, Italy}
\altaffiltext{7}{INAF, Osservatorio Astronomico di Torino, Via
  Osservatorio 20, I-10025 Pino Torinese, Italy}
\altaffiltext{8}{CRESST and NASA Goddard Space Flight Center,
  Greenbelt, MD, USA}
\altaffiltext{9}{Universities Space Research Association, Columbia,
  MD, USA}
% ...
\altaffiltext{*}{Email: \texttt{stefano@ifc.inaf.it}}

 \begin{abstract}
   We report on eighteen months of multiwavelength observations
   of the blazar 3C~454.3 ({\it Crazy Diamond}) carried out
   in the period July 2007 - January 2009. In particular,
   we show the results of the \agile{} campaigns
   which took place on May--June 2008, July--August 2008, 
   and October 2008 - January 2009. 
   During the May 2008 - January 2009 period,
   the source average flux was highly variable,
   with a clear fading trend towards the end of the period, from an average \gray flux 
   $ F_{\rm E>100MeV} \ga 200 \times 10^{-8}$\,\phcmsec\, in May--June 2008,
   to $ F_{\rm E>100MeV} \sim 80 \times 10^{-8}$\,\phcmsec\, in
   October 2008 - January 2009. 
   The average \gray spectrum between 100~MeV and 1~GeV
   can be fit by a simple power law, showing a moderate softening
   (from $\Gamma_{\rm GRID} \sim 2.0$  to $\Gamma_{\rm GRID} \sim 2.2$)
   towards the end of the observing campaign.
   Only $3\,\sigma$ upper 
   limits can be derived in the 20--60~keV energy band with Super-AGILE, because 
   the source was considerably off-axis during the whole time period.

   In July--August 2007 and May--June 2008, 3C~454.3 was monitored by 
   RXTE. 
   The \rxte{}/PCA light curve in the 3--20 keV energy band shows
   variability correlated with the \gray one.
   The \rxte{}/PCA average flux during the two time periods 
   is $F_{\rm 3-20 keV} = 8.4 \times 10^{-11}$\,\ergcmsec, and 
   $F_{\rm 3-20 keV} = 4.5 \times 10^{-11}$\,\ergcmsec, respectively, while 
   the spectrum (a power-law with photon index 
   $\Gamma_{\rm PCA} = 1.65 \pm 0.02)$ does 
   not show any significant variability. Consistent results are
   obtained with the analysis of the \rxte{}/HEXTE quasi-simultaneous data.

   We also carried out simultaneous \swi{} observations during 
   all \agile{} campaigns. \swi{}/XRT detected \source{} with an observed 
   flux in the 2--10~keV energy band in the range
   $(0.9-7.5) \times 10^{-11}$\,\ergcmsec{} and a photon index
   in the range $\Gamma_{\rm XRT} = 1.33-2.04$. In the 15--150~keV
   energy band, when detected, the source has an average flux of about
   5~mCrab. 

   GASP-WEBT monitored \source{} during the whole 2007--2008 period in the
   radio, millimeter, near-IR, and optical bands. The observations show an extremely
   variable behavior at all frequencies, with flux peaks almost
   simultaneous with those at higher energies.
   A correlation analysis between the optical and the \gray fluxes
   shows that the $\gamma$-optical correlation occurs with a time 
   lag of $\tau=-0.4^{+0.6}_{-0.8}$ days, consistent with
   previous findings for this source.

   An analysis of 15~GHz and 43~GHz VLBI core radio flux observations in the period
   2007 July - 2009 February shows an increasing trend of the core
   radio flux, anti-correlated with the higher frequency data,
   allowing us to derive
   the value of the source magnetic field. 

   Finally, the modeling of the broad-band spectral energy distributions
   (SEDs) for the still unpublished data, and the behavior of the
   long-term light curves in different energy bands, 
   allow us to compare the jet properties during different emission
   states, and to study the geometrical properties of the jet on a
   time-span longer than one year.
\end{abstract}

%
%
       %%%%%%%%%%%%%%%%%%%%%%%%%%%%%%%%%%%%%%%%%%%%%%%%%%%%%%%%%%%%%%%%%%%%
                       \section{Introduction} \label{3c454:intro}
       %%%%%%%%%%%%%%%%%%%%%%%%%%%%%%%%%%%%%%%%%%%%%%%%%%%%%%%%%%%%%%%%%%%%
%
%
Among active galactic nuclei (AGNs), blazars show intense and variable
\gray emission above 100~MeV \citep{Hartman1999:3eg}, with variability timescales
as short as a few days, or a few weeks.

Blazars emit across several decades of energy, from the radio to the TeV
energy band, and their spectral energy distributions (SEDs) typically show two
distinct humps. The first peak occurs in
the IR/Optical band in the Flat Spectrum Radio Quasars (FSRQs) and in
the Low-energy peaked BL Lacs (LBLs), and at UV/X-rays in the 
High-energy peaked BL Lacs (HBLs). The second hump peaks at MeV--GeV 
and TeV energies in FSRQs/LBLs and in HBLs, respectively.
In the framework of leptonic models, the first peak is commonly 
interpreted as synchrotron radiation
from high-energy electrons in a relativistic jet, while the second peak is
interpreted as inverse Compton (IC) scattering of soft seed photons by
the same relativistic electrons.
A recent review of the blazar emission mechanisms and energetics
is given in \cite{Celotti2008:blazar:jet}.
Alternatively, the blazar SED can be explained in the framework of the
hadronic models, where the relativistic protons in the jet are the primary 
accelerated particles, emitting \gray radiation by means of photo-pair and 
photo-pion production (see \citealt{Mucke2001:hadronic,Mucke2003:hadronic} 
for a review on hadronic models).

Since the launch of \agile{}, the FSRQ \source{} (PKS~2251$+$158; $z=0.859$) became 
one of the most active sources in the \gray sky. Its
very high \gray flux (well above $100 \times 10^{-8}$\,\phcmsec{} for
$E>100$~MeV), its flux variability (on a time scale of 1 or 2 days), and
the fact that it was always detected during any \agile{} pointing, made
it earn the nickname of {\it Crazy Diamond}: \source{} is now playing 
the same role as 3C~279 had for \egret{} \citep[e.g., ][]
{Hartman2001:3C279:multiwave,Hartman2001:3C279:gammaopt}.

Multiwavelength studies of variable \gray blazars are crucial in order
to understand the physical processes responsible for the emission
along the whole spectrum.
Since the detection of the exceptional 2005 outburst (see 
\citealt{Giommi2006:3C454_Swift,Fuhrmann2006:3C454_REM,Pian2006:3C454_Integral}), 
several monitoring campaigns were carried out to follow the source multi frequency behavior 
\citep{vil06,vil07,rai07,rai08a,rai08b}. 
In mid July 2007, 3C~454.3 underwent a new optical brightening, 
which triggered observations at all frequencies. 
\agile{} performed a target of opportunity (ToO)
re-pointing towards the source and detected it in a very high \gray state 
\citep[][hereafter V08]{Vercellone2008:3C454_ApJ}. 
In November and December 2007, \agile{} detected high \gray activity from
\source{}, triggering multiwavelength ToO campaigns, whose results
are reported in \citet[][hereafter Paper~I]{Vercellone2008:3c454:ApJ_P1},
in \citet[][hereafter Paper~II]{Donnarumma2009:3c454:subm}, and in 
\cite{Anderhub2009:AA:3C454}, respectively.
Paper~I and Paper~II demonstrated that 
to fit the simultaneous broad-band SEDs from radio to \gray data, 
inverse Compton (IC) scattering
of external photons from the broad line region (BLR) off the relativistic 
electrons in the jet was required.
In an earlier work based on the \cite{Vercellone2007:atel1160}
preliminary flux estimate of the July 2007 flare,
\cite{Ghisellini2007:3C454:SED} made a comparison between the
\source{} SEDs in 2000 (EGRET data), 2005 (optical and X-ray flare),
and 2007 (\agile{} \gray flare), discussing the role of the bulk Lorentz
factor $\Gamma$ (associated with the emitting source compactness)
during the different epochs.

Moreover, the results of correlation analysis performed in Paper~I 
was consistent with no time-lags between 
the \gray and the optical flux variations. Such a result was recently
confirmed by \citet{Bonning2008:3C454_apj} who correlated optical, UV, 
X-ray and \gray\footnote{The \gray data are taken from the \glast{}/LAT 
monitored source list light curves available at
\texttt{http://fermi.gsfc.nasa.gov/ssc/data/access/}} data.
In a very recent paper, \cite{Abdo2009:3C454} show the results
of the first three months of \glast{} observations of \source{},
from 2008 July to October.
They present for the first time the signature of a spectral break
above a few GeV, interpreted as a possible break in the energy distribution 
of the emitting particles.

In this paper (Paper~III) we present both a re-analysis of the \agile{}
published data collected during the period July 2007 - December 2007, and
the results of multiwavelength campaigns
on \source{} during a long-lasting \gray activity period between
2008 May 10 and  2009 January 12. 
In particular, we show the results of the \agile{} campaigns
which took place on May--June 2008 (mj08), July--August 2008 (ja08), 
and October 2008 - January 2009 (oj09). 
Preliminary \gray results were distributed in
\citet{Donnarumma2008:ATel1545,Vittorini2008:ATel1581,Gasparrini2008:ATel1592,
Pittori2008:ATel1634}, while radio-to-optical data
were published in \citet{Villata2009:3C454:GASP:accep}.

The paper is organized as follows: in Sect.~\ref{3c454:grid} through~\ref{3c454:radio:vlbi}
we present the \agile{},
\swi{}, {\it Rossi} X-ray Timing Explorer (\rxte{}), 
GLAST-AGILE Support Program within the Whole Earth Blazar Telescope
(GASP-\webt{}) and radio VLBI data analysis and results;
in Sect.~\ref{3c454:monitoring} we present the simultaneous multiwavelength light-curves.
In Sect.~\ref{3c454:disc} and~\ref{3c454:conc} we discuss the results and draw our conclusions.
Throughout this paper the quoted uncertainties are given at the
$1\sigma$ level, unless otherwise stated, and we adopted a $\Lambda$-CDM cosmology with the
following values for the cosmological parameters: $h = 0.71$, $\Omega_{m}
= 0.27$, and $\Omega_{\Lambda} = 0.73$.
%
%

       %%%%%%%%%%%%%%%%%%%%%%%%%%%%%%%%%%%%%%%%%%%%%%%%%%%%%%%%%%%%%%%%%%%%
                       \section{\agile{} data} \label{3c454:grid}
       %%%%%%%%%%%%%%%%%%%%%%%%%%%%%%%%%%%%%%%%%%%%%%%%%%%%%%%%%%%%%%%%%%%%
%
%
%%%%%--------------------------------------------
     \subsection{Data Reduction and Analysis} \label{3c454:grid:analysis}
%%%%%--------------------------------------------
%
The AGILE satellite \citep{Tavani2008_agile_nima,Tavani2009:Missione},
a mission of the Italian Space Agency (ASI) devoted to high-energy
astrophysics, is currently the only space mission capable of
observing cosmic sources simultaneously in the energy bands
18--60~keV and 30~MeV--50~GeV. 

The AGILE scientific instrument combines four
active detectors yielding broad-band coverage from hard X-rays
to gamma-rays:
a Silicon Tracker~\citep[ST;][30~MeV--50~GeV]{Prest2003:agile_st},
a co-aligned coded-mask hard X-ray imager, Super--AGILE
\citep[SA;][18--60~keV]{Feroci2007:agile_sa}, a non-imaging CsI
Mini--Calorimeter~\citep[MCAL;][0.3--100~MeV]{Labanti2009:agile_mcal},
and a segmented Anti-Coincidence System~\citep[ACS;][]{Perotti2006:agile_ac}.
Gamma-ray detection is obtained by the combination of ST,
MCAL and ACS; these three detectors form the AGILE Gamma-Ray Imaging
Detector (GRID).

Level--1 AGILE-GRID data were analyzed using the AGILE Standard Analysis
Pipeline (see~V08 for a detailed description of
the AGILE data reduction). Since \source{} was observed at high off-axis 
angle due to the satellite solar panel constraints, 
an ad-hoc \gray analysis was performed. We used \gray events 
filtered by means of the \texttt{FM3.119$\_$2a} \agile{} Filter pipeline.
Counts, exposure, and Galactic background \gray maps were created with
a bin-size of $0.\!\!^{\circ}25 \times 0.\!\!^{\circ}25$\,,
for $E \ge 100$\,MeV. Since the source was observed up to $40^{\circ}$ off-axis,
all the maps were generated including all events collected up to
$60^{\circ}$ off-axis.
We rejected all \gray events whose reconstructed
directions form angles with the satellite-Earth vector smaller
than $85^{\circ}$,
reducing the \gray Earth albedo contamination
by excluding regions within $\sim 15^{\circ}$ from the
Earth limb.
We used the version (\texttt{BUILD-16}) of the Calibration files
(\texttt{I0006}),
which are publicly available at the ASI Science Data Centre
(ASDC) site\footnote{\texttt{http://agile.asdc.asi.it}},
and of the \gray diffuse emission model \citep{Giuliani2004:diff_model}.
We ran the AGILE Source Location task in order
to derive the most accurate location of the source. Then,
we ran the AGILE Maximum Likelihood Analysis (\texttt{ALIKE}) using
a radius of analysis of 10$^{\circ}$,
and the best guess position derived
in the first step. The particular choice of the analysis radius
is dictated to avoid any possible contamination from very off-axis residual
particle events.
%
%
%%%%%--------------------------------------------
     \subsection{GRID Results} \label{3c454:grid:results}
%%%%%--------------------------------------------
%
Table~\ref{3c454:tab:grid:obs_log} shows the \agile{}/GRID observation
log during the different time periods. We have re-analysed all the
\agile{} data already published in V08, Paper~I, and
Paper-II, respectively, according to the procedure described in 
Section~\ref{3c454:grid:analysis}. The results are discussed in 
Section~\ref{3c454:monitoring}.
%
%        TABLE  - GRID OBSERVATION LOG
%
\begin{deluxetable}{ccccc} 	
  \tablecolumns{4}
  \tabletypesize{\normalsize}
  \tablecaption{\agile{}/GRID observation log.\label{3c454:tab:grid:obs_log}}
  \tablewidth{0pt}
  \tablehead{
    \colhead{Epoch} & \colhead{Start Time} & \colhead{End Time}    & \colhead{Exposure} \\
    \colhead{}      & \colhead{(UTC)}      & \colhead{(UTC)}       & \colhead{(Ms)} }
  \startdata
    1   & 2007-07-24 14:30   &   2007-07-30 11:40  &  $0.22$  \\
    2   & 2007-11-10 12:16   &   2007-12-01 11:38  &  $0.64$  \\
    3   & 2007-12-01 11:39   &   2007-12-16 12:09  &  $0.56$  \\
    4   & 2008-05-10 11:00   &   2008-06-09 15:20  &  $1.03$  \\
    5,6 & 2008-06-15 10:46   &   2008-06-30 11:14  &  $0.54$  \\
    7   & 2008-07-25 19:57   &   2008-08-14 21:08  &  $0.70$  \\
    8   & 2008-10-17 12:51   &   2009-01-12 14:30  &  $2.86$  \\
  \enddata
\end{deluxetable}
In the following paragraphs we report the detailed results of the
still unpublished \agile{} data.

%%%........
     \subsubsection{May--June 2008} \label{3c454:grid:results:mj08}
%%%........
The \agile{} campaign was split into two different periods,
May 10--June 9 (P1) and June 15--30 (P2) because of a ToO re-pointing 
towards W~Comae. The total on-source exposure is 1.03 (P1) + 0.54 (P2)~Ms.
\source{} was detected, during P1 and P2, at a $25.6\sigma$ and $16.3\sigma$
level with an average flux of
$ F_{\rm E>100MeV}^{\rm P1} = (218 \pm 12) \times 10^{-8}$\,\phcmsec,
and $ F_{\rm E>100MeV}^{\rm P2} = (198 \pm 17) \times 10^{-8}$\,\phcmsec\,,
respectively, as derived from the \agile{} Maximum Likelihood Code analysis.

\begin{figure}[ht]
\resizebox{\hsize}{!}{\includegraphics[angle=-90]{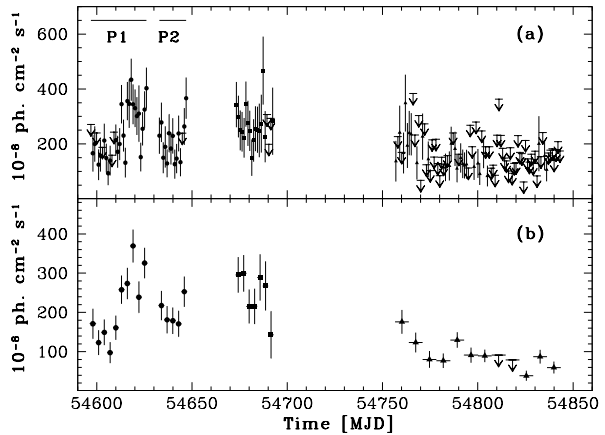}}
  \caption{Panel (a): AGILE/GRID \gray light curve at
       $\approx 1$-day resolution for E$>$100~MeV in units of
      $10^{-8}$\,\phcmsec\, during the period May 2008 - January
      2009. The downward arrows represent $2\sigma$ upper-limits.
      Panel (b): the same as in panel (a), but with a time bin of
      three (filled circles and squares), and seven (filled triangles)
      days, respectively.
  }
  \label{3c454:figure1}
\end{figure}

Figure~\ref{3c454:figure1}, filled circles in panel (a), 
shows the \gray light-curve
at 1-day resolution for photons above 100~MeV.
%
%----------------- FIGURE 1 : LIGHT CURVES --------------------
%-------------------------------------------------------------
%

%
We note that, particularly at the beginning of the campaign,
\source{} was not always detected on a day-by-day timescale. On
MJD $\sim$~54610 the source began to be detected at a $3\sigma$ level
almost continuously; this clearly indicates the onset of a \gray flaring 
activity.
The average \gray flux as well as the daily values were
derived according to the \gray analysis procedure described in
\citet{Mattox1993:1633}.
First, the entire period was analyzed to determine the diffuse
gas parameters and then the source flux density was estimated
independently for each of the eighteen 1-day periods with the diffuse
parameters fixed at the values previously obtained.

Figure~\ref{3c454:figure1}, panel (b), shows the same
\agile{}/GRID data binned on a time scale of three days. The
light curve clearly shows a strong degree of variability, with
a dynamic range of about four in about two weeks.

\begin{figure}[ht]
\resizebox{\hsize}{!}{\includegraphics[angle=0]{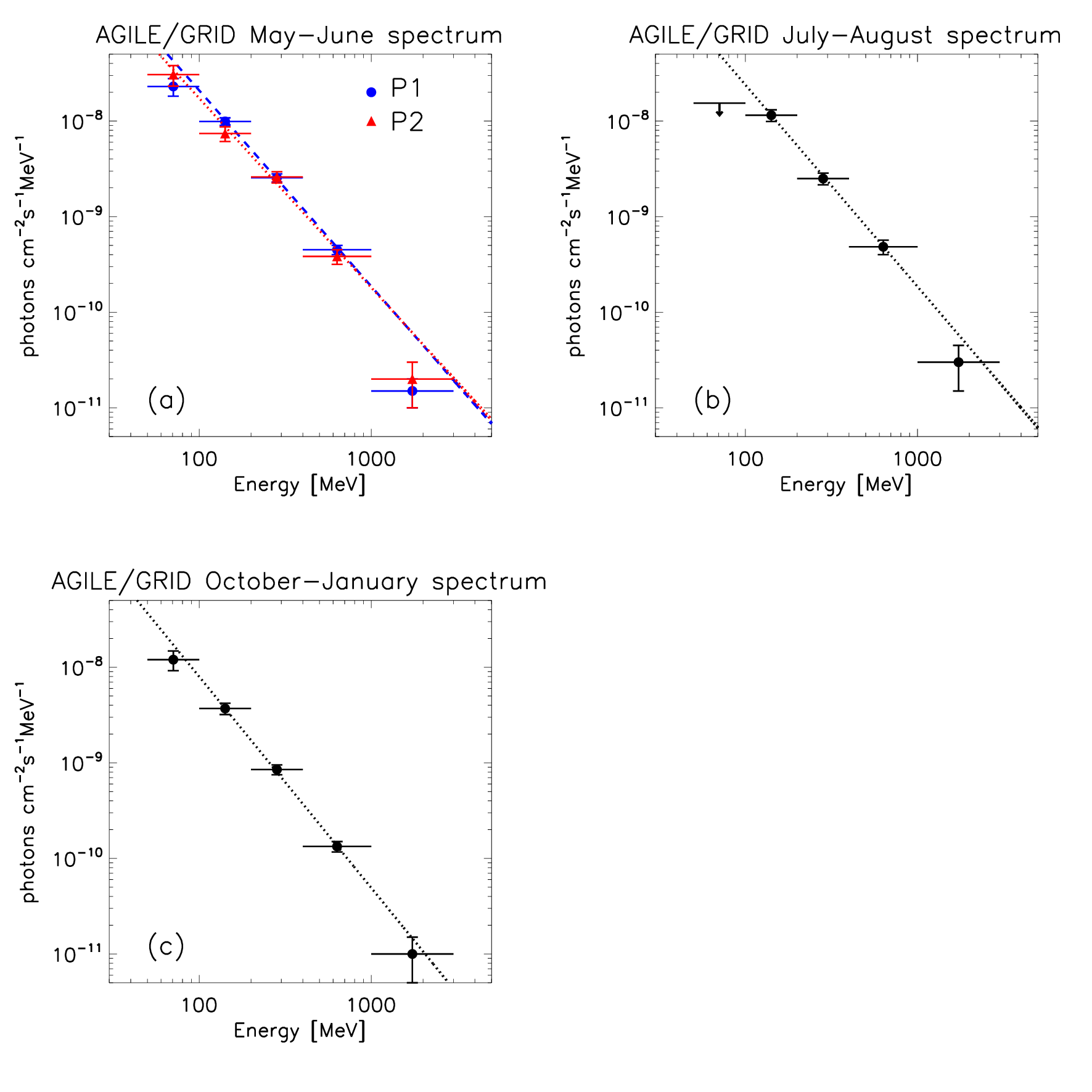}}
  \caption{Panel (a): AGILE/GRID average \gray spectrum 
    for periods P1 and P2. 
    The blue-dashed and the red-dotted lines represent the best--fit 
    power law models for P1 and P2, respectively.
    Panel (b) and Panel (c) show the average \gray spectra during the
    periods July--August 2008 and October 2008 - January 2009,
    respectively. 
    In the three panels only three energy bins were
    considered for the spectral fitting: $100 < {\rm E} < 200$~MeV, $200 < {\rm E} < 400$~MeV,
    $400 < {\rm E} < 1000$~MeV (see text for details).
    [{\it See the electronic edition of the Journal for a color version of
      this figure.}]
  }
  \label{3c454:figure2}
\end{figure}

Figure~\ref{3c454:figure2}, panel (a), shows the average \gray spectra
extracted over the observing periods P1 and P2.
%
%----------------- FIGURE 2 : GRID SPECTRUM --------------------
%-------------------------------------------------------------
%

%
Each average spectrum was obtained by computing the \gray flux in
five energy bins over each period:
$50 < {\rm E} < 100$~MeV,
$100 < {\rm E} < 200$~MeV, $200 < {\rm E} < 400$~MeV,
$400 < {\rm E} < 1000$~MeV, and $1000 < {\rm E} < 3000$~MeV.
We note that the current instrument response is accurately calibrated
in the energy band 100~MeV--1~GeV, and that the flux
above 1~GeV is underestimated by a factor of about 2. For those
reasons, we fit the data by means of a simple power law model and restricted
our fit to the 100~MeV--1~GeV energy range, obtaining (in units of 
${\rm photons\hspace{2mm}cm^{-2} ~s^{-1} ~MeV^{-1}}$):
\begin{equation}\label{3c454:equ:deffluxp1}
  F^{\rm P1}(E) =
  2.63 \times 10^{-4} \times
  \left( \frac{{\rm E}}{1\, {\rm MeV}}\right)^{-(2.05 \pm 0.10)}\,,
\end{equation}
\begin{equation}\label{3c454:equ:deffluxp2}
  F^{\rm P2}(E) =
  1.58 \times 10^{-4} \times
  \left( \frac{{\rm E}}{1\, {\rm MeV}}\right)^{-(1.98 \pm 0.16)}\,.
\end{equation}
The different energy range, and, above all, the different time period,
could explain the different value of the \agile{} ($2.05 \pm 0.10$,
and $1.98 \pm 0.16$) 
and \glast{}-LAT ($2.27 \pm 0.03$\,, pre-break, \citealt{Abdo2009:3C454}) \gray photon indices.
%

%%%........
     \subsubsection{July--August 2008} \label{3c454:grid:results:ja08}
%%%........
The \agile{} campaign started immediately after the \glast{}/LAT detection
of a very high \gray activity in the period 2008 July 10--21, 
\citep{Tosti2008:ATel1628}, which reached, on July 10, a \gray flux
of $ F_{\rm E>100MeV}^{\rm Fermi} = 1200 \times 10^{-8}$\,\phcmsec\,
\citep{Abdo2009:3C454}.
The \agile{} observations covered the period from 2008-07-25 19:57 UT to
2008-08-14 21:08 UT, for a total on-source exposure of about 0.71~Ms.
\source{} was detected at a $17.5\sigma$
level with an average flux of $ F_{\rm E>100MeV}^{\rm ja08} = (255 \pm 21) 
\times 10^{-8}$\,\phcmsec, as derived from the \agile{} Maximum Likelihood 
Code analysis.

Figure~\ref{3c454:figure1}, filled squares in panel (a) and in panel (b),
shows the \gray light-curve
at 1-day and at 3-day resolution, respectively, for photons above 100~MeV.
The average \gray flux as well as the daily values were 
derived according to the procedure described in \S~\ref{3c454:grid:results:mj08}\,.
Contrary to the May--June data, the July--August light curve does
not show any clear sign of variability.

Figure~\ref{3c454:figure2}, panel (b), shows the average \gray spectrum
derived over the entire observing period.
The average spectrum was obtained by computing the \gray flux in the same
way as in \S~\ref{3c454:grid:results:mj08}.
We fit the data by means of a simple power law model and restricted
our fit to the most reliable energy range (100~MeV--1~GeV):
\begin{equation}\label{3c454:equ:deffluxja08}
  F^{\rm ja08}(E) =
  3.96 \times 10^{-4} \times
  \left( \frac{{\rm E}}{1\, {\rm MeV}}\right)^{-(2.11 \pm 0.14)}\,,
\end{equation}
in units of ${\rm photons\hspace{2mm}cm^{-2} ~s^{-1} ~MeV^{-1}}$.
During this period, which partially overlaps with the \glast{} one, the \agile{} 
\gray photon index is, within the statistical errors, in agreement with the 
\glast{}-LAT result.
%

%%%........
  \subsubsection{October 2008 - January 2009} \label{3c454:grid:results:oj09}
%%%........
The \agile{} observations covered the period from 2008-10-17 12:51 UT to
2009-01-12 14:30 UT, for a total on-source exposure of about 2.86~Ms.
\source{} was detected at a $17.9\sigma$
level with an average flux of $ F_{\rm E>100MeV}^{\rm oj09} = (77 \pm 5) 
\times 10^{-8}$\,\phcmsec, as derived from the \agile{} Maximum Likelihood 
Code analysis.

Figure~\ref{3c454:figure1}, filled triangles in panel (a),
shows the \gray light-curve
at 1-day resolution for photons above 100~MeV.
The average \gray flux as well as the daily values were 
derived according to the procedure described in \S~\ref{3c454:grid:results:mj08}.
The light curve does not show any clear trend, partly because of a dominant
fraction of upper limits in the data. For this reason, we decided to
rebin the light curve on a time scale of one week. The resulting light
curve is shown in Figure~\ref{3c454:figure1}, filled triangles in panel (b).
On this time scale, a clear trend is present, with the source dimming
as a function of time, with a dynamic range of about a factor of two. 

Figure~\ref{3c454:figure2}, panel (c), shows the average \gray spectrum
derived over the entire observing period.
The average spectrum was obtained by computing the \gray flux in the same
way as in \S~\ref{3c454:grid:results:mj08}.
We fit the data by means of a simple power law model and restricted
our fit to the most reliable energy range (100~MeV--1~GeV):
\begin{equation}\label{3c454:equ:deffluxoj09}
  F^{\rm oj09}(E) =
  2.10 \times 10^{-4} \times
  \left( \frac{{\rm E}}{1\, {\rm MeV}}\right)^{-(2.21 \pm 0.13)}\,,
\end{equation}
in units of ${\rm photons\hspace{2mm}cm^{-2} ~s^{-1} ~MeV^{-1}}$

%
%
%%%%%--------------------------------------------
     \subsection{Super-AGILE Results} \label{3c454:sa:results}
%%%%%--------------------------------------------
%
%
During the various \agile{} pointings, \source{} was located substantially
off-axis in the Super--AGILE field of view (FoV). For this reason, only $3\sigma$ upper limits
can be derived in the 20--60~keV energy band during the \agile{}/GRID observations.
Table~\ref{3c454:tab:sa} summarizes the Super--AGILE observations and
their results.
%
%        TABLE  - SA OBSERVATION LOG
%
\begin{deluxetable*}{cccccc} 	
  \tablecolumns{4}
  \tabletypesize{\normalsize}
  \tablecaption{Super--AGILE observation results.\label{3c454:tab:sa}}
  \tablewidth{0pt}
  \tablehead{
   \colhead{Start Time} & \colhead{End Time} & \colhead{$\theta_{\rm X}$}  
       & \colhead{$\theta_{\rm Z}$}  & \colhead{Exposure} & \colhead{$F_{\rm 20-60\,keV}$} \\
   \colhead{(UTC)}      & \colhead{(UTC)}    & \colhead{(Deg.)} 
       & \colhead{(Deg.)}   & \colhead{(ks)}  & \colhead{(mCrab)}}
  \startdata
    2008-05-31 10:18  &  2008-06-09 13:38  & $-23.0$  & $+06.0$ & $380$ & $<16$\\
    2008-06-15 14:11  &  2008-06-21 12:59  & $-36.0$  & $+08.0$ & $270$ & $<18$\\
    2008-07-25 21:39  &  2008-08-02 23:29  & $+03.4$  & $-42.0$ & $345$ & $<18$\\
    2008-10-17 18:47  &  2008-10-29 23:12  & $-00.8$  & $-45.0$ & $460$ & $<21$\\
  \enddata
\end{deluxetable*}
%

       %%%%%%%%%%%%%%%%%%%%%%%%%%%%%%%%%%%%%%%%%%%%%%%%%%%%%%%%%%%%%%%%%%%%
                       \section{\swi{} data} \label{3c454:swift}
       %%%%%%%%%%%%%%%%%%%%%%%%%%%%%%%%%%%%%%%%%%%%%%%%%%%%%%%%%%%%%%%%%%%%
%
%
%%%%%--------------------------------------------
     \subsection{Data Reduction and Analysis} \label{3c454:swift:analysis}
%%%%%--------------------------------------------

\swi{} pointed observations \citep{Gehrels2004:swift} were performed from 2007-07-26
to 2009-01-01. These observations were obtained both by means of several
dedicated ToOs (PI S.\ Vercellone) and by activating \swi{} Cycle-3 
(Obs. ID 00031018 PI A.W.\ Chen) and Cycle-4 Proposals 
(Obs. ID 00031216, PI S.\ Vercellone).
A long-lasting monitoring program (P.Is. L.Fuhrmann and S. Vercellone)
covers the period July--October 2008. 
%

%%%%%--------------------------------------------
\subsubsection{\swi{}/XRT} \label{3c454:swift:analysis:XRT}
%%%%%--------------------------------------------

Table~\ref{3c454:tab:xrt:obs_log} summarizes the \swi{}/XRT observations.
The XRT data were processed with standard procedures
({\tt xrtpipeline} v0.12.1),
adopting the standard filtering and screening criteria, and using FTOOLS
in the {\tt Heasoft} package (v.6.6.1).
%
%        TABLE  - XRT OBSERVATION LOG
%
 \begin{deluxetable*}{ 	
l	l	l	l	r	}
  \tablewidth{0pc} 	      	 
  \tablecaption{\swi{}/XRT observation log. \label{3c454:tab:xrt:obs_log}} 
  \tablehead{	
\colhead{Sequence}    & \colhead{Obs.}    & \colhead{Start time  (UT)}    & \colhead{End time   (UT)}    & \colhead{Exp.\tablenotemark{a}} \\
\colhead{}    & \colhead{Mode}    & \colhead{(yyyy-mm-dd hh:mm:ss)}    & \colhead{(yyyy-mm-dd hh:mm:ss)}    & \colhead{s} \\
\colhead{(1)} & \colhead{(2)} & \colhead{(3)} & \colhead{(4)} & \colhead{(5)} 
}
  \startdata
00035030013	&  PC	 & 2007-07-26 00:55:44	  & 2007-07-26 01:13:58  &  1073    \\
00035030014	&  PC	 & 2007-07-28 07:26:46	  & 2007-07-28 10:44:55  &  817     \\
00035030015	&  PC	 & 2007-07-30 10:59:09	  & 2007-07-30 14:16:56  &  897     \\
00035030016	&  PC	 & 2007-08-01 11:05:10	  & 2007-08-01 11:07:58  &  168     \\
00035030017	&  WT	 & 2007-08-01 09:33:17	  & 2007-08-01 13:06:59  &  3903    \\
    \enddata  
     \tablenotetext{1}{NOTE.--Table~\ref{3c454:tab:xrt:obs_log} is published in its entirety in the
    electronic edition of the {\it Astrophysical Journal}. 
    A portion is shown here for guidance regarding its form and
    content. }
    \tablenotetext{a}{The exposure time is spread over several snapshots during 
      each observations} 
   \end{deluxetable*}  

The source count rate was variable during the campaigns,
ranging from 0.26 to 1.8 counts\,s$^{-1}$\,. For this reason,
we considered both photon counting (PC) and windowed timing (WT) data,
and further selected XRT event grades 0--12 and 0--2 for the PC and WT
events, respectively \citep{Burrows2005:grades}.
Several \swi{}/XRT observations showed an average count rate of $>0.5$ counts s$^{-1}$,
therefore in these cases pile-up correction was required for the PC data.
We extracted the source events from  an annular
region with an inner radius of 3 pixels (estimated
by means of  the PSF fitting technique) and an outer radius of 30 pixels
(1 pixel $\sim2\farcs36$). When the average count rate was $<0.5$
counts s$^{-1}$, we used the full 30-pixel radius region.

We also extracted background events within an
annular region centered on the source with radii of 110 and 116 pixels.
Ancillary response files were generated with {\tt xrtmkarf},
and account for different extraction regions, vignetting and
PSF corrections. We used the spectral redistribution matrices
v011 in the Calibration Database maintained by HEASARC.
\swi{}/XRT uncertainties are given at
90\% confidence level
for one interesting parameter (i.e., $\Delta \chi^2 =2.71$)
unless otherwise stated.

The \swi{}/XRT spectra were rebinned in order to have at least 
20 counts per energy bin. 
We fit the spectra with an absorbed power law model,
(\texttt{wabs*(powerlaw)} in \texttt{XSPEC 11.3.2}).
The Galactic absorption was fixed to the value of
$N_{\rm H}^{\rm Gal} = 1.34 \times 10^{21}$\,cm$^{-2}$,
as obtained by \cite{vil06} by means of a deep {\it Chandra}
observation.
We note the adopted value is consistent with the mean of the 
distribution of the $N_{\rm H}$ values obtained by fitting 
the spectra with an absorbed power law model and free absorption.

%
%%%%%--------------------------------------------
\subsubsection{\swi{}/UVOT} \label{3c454:swift:analysis:UVOT}
%%%%%--------------------------------------------
%

%
The UVOT data analysis was performed using the {\tt uvotimsum} and
{\tt uvotsource} tasks included in the {\tt FTOOLS} software package
(HEASOFT v6.6.1).
The latter task calculates the magnitudes through aperture photometry within
a circular region and applies specific corrections due to the detector
characteristics.
Source counts were extracted from a circular region with a 5 arcsec radius.
The background was extracted from source-free circular regions in the
source surroundings. The reported magnitudes 
are on the UVOT photometric system described in \citet{Poole2008:UVOT}, and are not corrected
for Galactic extinction.

%
%%%%%--------------------------------------------
\subsubsection{\swi{}/BAT} \label{3c454:swift:analysis:BAT}
%%%%%--------------------------------------------
We analyzed \swi{}/BAT Survey data in order to study the hard X-ray
emission of \source{} and to investigate its evolution as a function of time.

We produced a light curve for the source at a 16-d binning
using the procedures described in
\citet[][and references therein; also see
\footnote{\texttt{http://swift.gsfc.nasa.gov/docs/swift/results/transients/Transient\_synopsis.html}}]{Krimm2006_atel_BTM,Krimm2008_HEAD_BTM}.

%
%%%%%--------------------------------------------
     \subsection{Results} \label{3c454:swift:results}
%%%%%--------------------------------------------
%

\begin{figure}[ht]
\resizebox{\hsize}{!}{\includegraphics[angle=0]{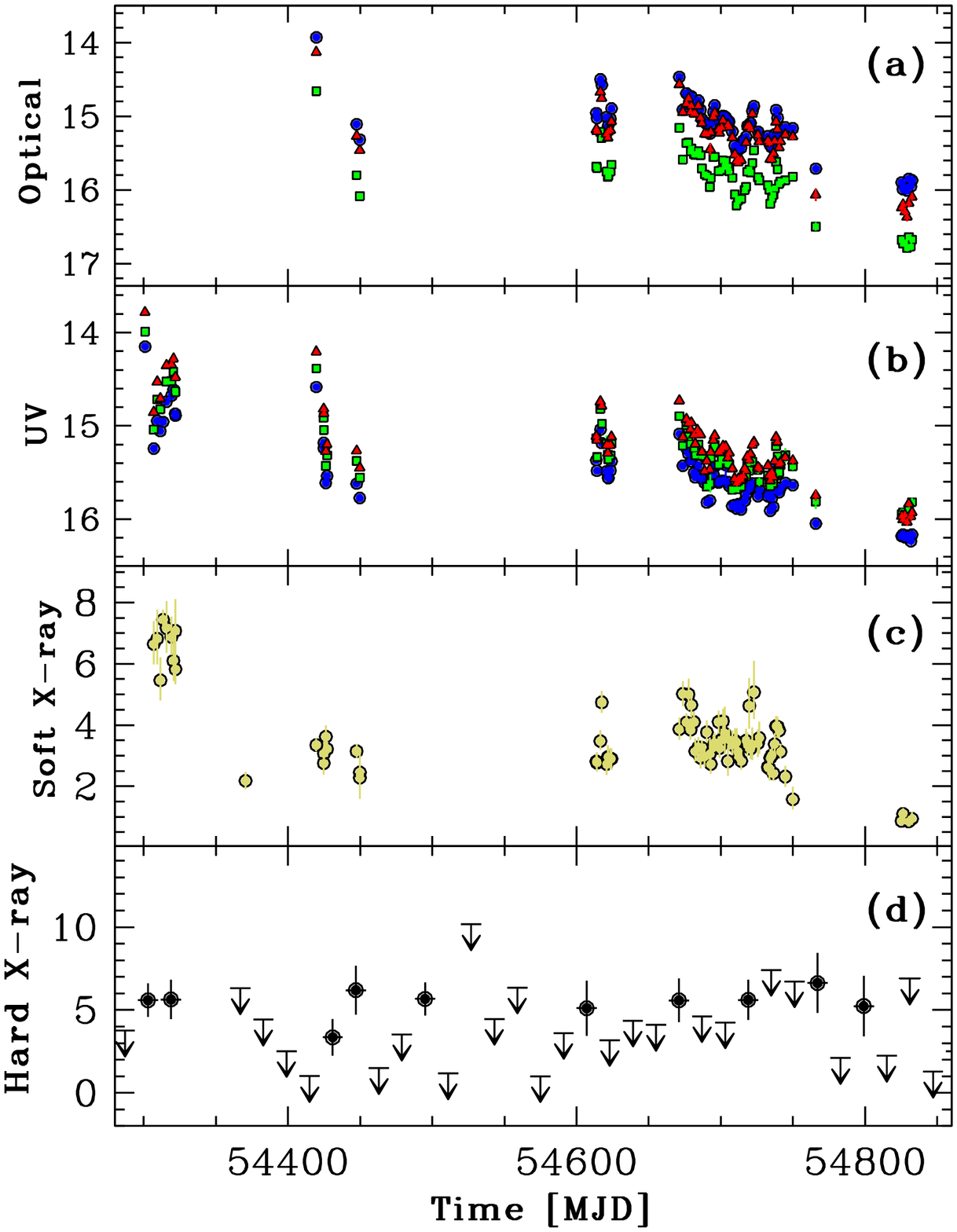}}
  \caption{
    {\it Panel (a)}: \swi{}/UVOT light curves (observed magnitudes) in the $V$ 
    (red triangles), $B$ (green quares), and $U$ (blue circles).
    {\it Panel (b)}: \swi{}/UVOT light curves (observed magnitudes) in the $W1$ 
    (red triangles), $M2$ (green quares), and $W2$ (blue circles).
    {\it Panel (c)}: \swi{}/XRT light curve (observed fluxes) in the 2-10~keV energy
    band and in units of $10^{-11}$ erg cm$^{-2}$s$^{-1}$.
    {\it Panel (d)}: \swi{}/BAT light curve in units of mCrab in the
    energy band $15-150$\,keV. Downward arrows
    mark $3\sigma$ upper limits.
  }
  \label{3c454:fig:swift:xrt_uvot_bat}
\end{figure}

Figure~\ref{3c454:fig:swift:xrt_uvot_bat}
shows the \swi{}/XRT fluxes in the 2--10~keV energy range, the \swi{}/UVOT 
observed magnitudes (in the $V$, $B$, $U$, $W1$, $M2$, and $W2$
bands), and the \swi{}/BAT fluxes in the 15--150~keV energy range
as a function of time for the whole observing period.
%
%----------------- FIGURE 3 : SWIFT UVOT + XRT + BAT --------------------
%-------------------------------------------------------------

In order to diminish the statistical uncertainties, we selected observations
with a number of degrees of freedom (dof)~$> 10$. 
We note that a common dimming trend is present both in the UV and in
the X-ray energy bands. 

As shown in Figure~\ref{3c454:fig:swift:xrt_uvot_bat}, panel (d),
the source has not been always detectable by \swi{}/BAT
throughout the considered period,
and in several time interval only $3\sigma$ upper limits
can be derived.

\begin{figure}[ht]
\resizebox{\hsize}{!}{\includegraphics[angle=0]{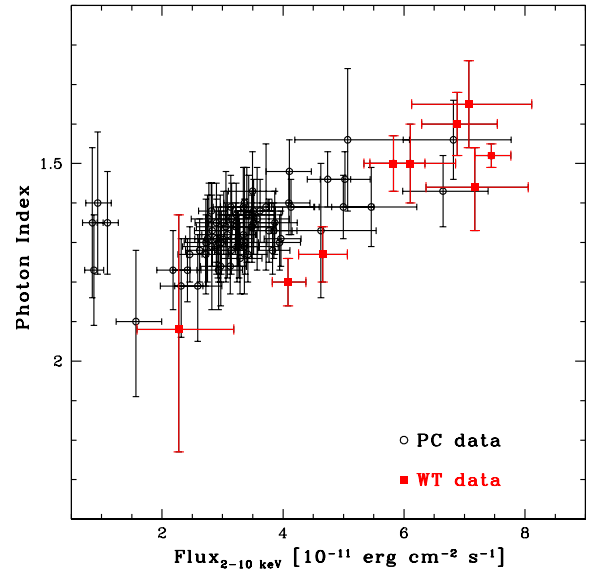}}
  \caption{\swi{}/XRT photon index as a function of the 2--10~keV
    flux. Red squares and black circles mark the \swi{}/XRT windowed
    timing (WT) and photon counting (PC) data, respectively.
  }
  \label{3c454:fig:swift:hister}
\end{figure}

Figure~\ref{3c454:fig:swift:hister} shows the \swi{}/XRT photon index
as a function
of the X-ray flux in the 2--10~keV energy band. Black circles and
red squares represent data acquired in PC and WT mode, respectively. 
WT data are not affected by pile-up at the observed count rate
($CR < 3$\,counts\,s$^{-1}$).
%
%----------------- FIGURE 4 : SWIFT  --------------------
%-------------------------------------------------------------
%
We investigated the possible presence of a spectral trend in the X-ray data. If we
consider WT data only, a ``harder-when-brighter'' trend seems to be
present. Fitting the data with a constant model, we can exclude this
model at the 99.9993\% level.
When analyzing the PC data only (as well as the sum of the PC and WT
data), this spectral trend vanishes, and a fit with a constant model
still holds. Nevertheless, if we exclude the points at fluxes
$F_{\rm 2-10\,keV} < 2 \times 10^{-11}$\,\ergcmsec a trend still
holds. These points at low fluxes could correspond to physically
different state of the source than the high fluxes one.

We also note that a deep and prolonged monitoring of \source{} at mid
and low X-ray states ($F_{\rm 2-10\,keV} \la 10^{-11}$\,\ergcmsec) will
be crucial to test the possible presence of a spectral trend. Our data-set
contains only four observations (90023002, 90023003, 90023006, and 90023008)
at this flux level, which  were acquired during the source low state
in December 2008.

%
%
       %%%%%%%%%%%%%%%%%%%%%%%%%%%%%%%%%%%%%%%%%%%%%%%%%%%%%%%%%%%%%%%%%%%%
              \section{\rxte{} data} \label{3c454:rxte}
       %%%%%%%%%%%%%%%%%%%%%%%%%%%%%%%%%%%%%%%%%%%%%%%%%%%%%%%%%%%%%%%%%%%%
%
%
%%%%%--------------------------------------------
     \subsection{Data Reduction and Analysis} \label{3c454:rxte:analysis}
%%%%%--------------------------------------------
%

The {\it Rossi X-ray Timing Explorer} ({RXTE}) satellite
observed \source{} in two epochs: from 2007-07-28 to 2007-08-04 and from
2008-05-30 to 2008-06-19. Here we report the analysis of the data
obtained both with the {\it Proportional Counter Array}
\citep[PCA,][]{Jahoda1996SPIE:RXTE},
which is sensitive in the 2--60~keV energy range,
and with the {\it High-Energy X-Ray Timing Experiment }
\citep[HEXTE,][]{Rothschild1998:HEXTE}, which is sensitive in
the 15--250~keV energy range.
\rxte{} data were collected by activating a Cycle 12 ToO proposal
(ID. 93150, PI A.W.~Chen).

The {PCA} is composed of 5 identical
units ({\it Proportional Counter Units}, PCUs), but during our
observations only part of them were used. Since PCU2 was the
only unit always on during our observations and it is the one which
is best calibrated, we report the results obtained from the
PCU2 data only.
The data were processed using the FTOOLS v6.4.1 and screened using
standard filtering criteria.
The net exposure times for the whole data-set in the first and second
epoch were 36.6~ks and 17.4~ks, respectively.

The background light curves and spectra for each observation were
produced using the model appropriate to faint sources.
We restricted our analysis to the 3--20 keV
energy range, in order to minimize the systematic errors due to
background subtraction and calibration of the {\it PCA} instrument.

\begin{figure}[ht]
\resizebox{\hsize}{!}{\includegraphics[angle=-90]{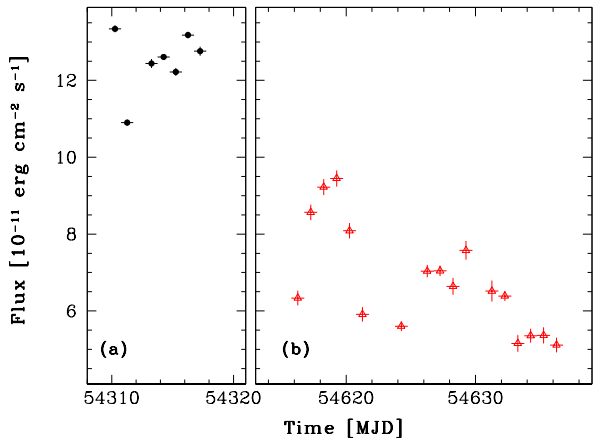}}
  \caption{Panel (a) and (b) show the \rxte{}/PCA light curve
    in the energy band 3-20 keV during the periods 
    2007 July 29 - August 5 (black points) and 2008 May 30 - June 19
    (red triangles), respectively.
    [{\it See the electronic edition of the Journal for a color version of
      this figure.}]
  }
  \label{3c454:fig:rxte:lc}
\end{figure}

Figure~\ref{3c454:fig:rxte:lc} shows the 3--20 keV light curve of 
the whole \rxte{}/PCA data-set.
%
%----------------- FIGURE 5 : RXTE LC --------------------
%-------------------------------------------------------------

Strong variability is observed when comparing the count rates of
different observations. Moreover, the average count rate during the second
epoch (panel (b)) is reduced. In order to investigate possible
changes in the spectral shape with time we extracted light curves in
two energy ranges (3--7 keV and 7--20 keV). Their hardness ratio
did not show any significant variation.

\begin{figure}[ht]
\resizebox{\hsize}{!}{\includegraphics[angle=-90]{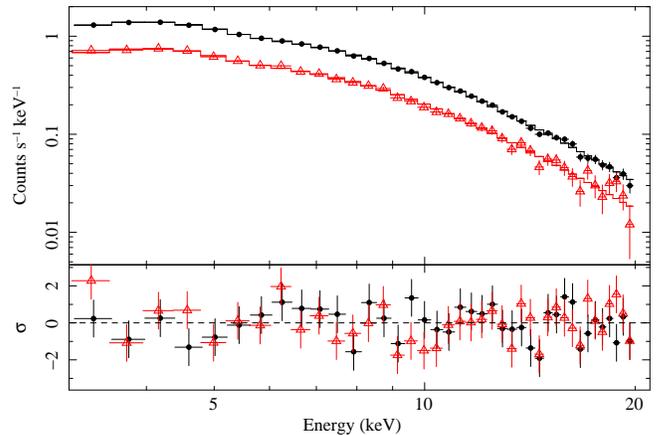}}
  \caption{\rxte{}/PCA average spectra for both
    periods, July 2007 (black points), and May-June 2008 (red triangles),
    respectively.
    [{\it See the electronic edition of the Journal for a color version of
      this figure.}]
  }
  \label{3c454:fig:rxte:spectra}
\end{figure}

A cumulative spectrum for the first and the second epoch was
extracted and simultaneously fitted with a power-law model corrected for
photoelectric absorption (\texttt{wabs*(powerlaw)} model in XSPEC), allowing only
the power-law normalization to assume a different value in the two
spectra. 
Figure~\ref{3c454:fig:rxte:spectra} shows the \rxte{}/PCA spectra for both
periods. 
A good fit ($\chi^{2}$= 68.3 for 76 degrees of freedom) 
was obtained with the following best-fit parameters
(errors are at the 90\% confidence level): 
photon index $\Gamma=  1.65\pm$0.02, and a flux
in the 3--20\,keV energy band $F_{\rm 3-20\,keV} = 8.4\times10^{-11} $ erg cm$^{-2}$ s$^{-1}$ 
and $F_{\rm 3-20\,keV} = 4.5\times10^{-11} $ erg cm$^{-2}$ s$^{-1}$ 
for the first and
second epoch spectrum, respectively. 
%
%----------------- FIGURE 6 : RXTE SPECTRA --------------------
%-------------------------------------------------------------

We note that the average \rxte{}/PCA flux during the \gray flare detected
in July 2007 was about a factor of two higher than the
flux detected about 10 months later. Moreover, during both the
July 2007 and the May-June 2008 campaigns, the hard X-ray flux
varied significantly, by about 50\%, on a time scale of about one week.

We also analyzed the data of the HEXTE.
Only the data from cluster 
B were analyzed, since the rocking system for background evaluation was 
disabled in the other instrument cluster. After a standard 
processing,\footnote{\texttt{http://heasarc.gsfc.nasa.gov/docs/xte/recipes/hexte.html}} 
we extracted an average spectrum from all the available data, for a
dead-time corrected exposure time of 18~ks. The source was detected up to $\sim 50$~keV and its 
spectrum can be fit well ($\chi^{2}$=15.1 for 19 d.o.f.) by a power-law 
model with a photon index of $1.6\pm 0.1$, in perfect agreement with the 
photon index derived from the PCA spectrum. Also, the normalization of the HEXTE 
spectrum is consistent with an high energy extrapolation of the 
time-averaged PCA spectrum: the observed flux in the 20--40~keV energy
band is (5$\pm$3)$\times$10$^{-11}$ erg cm$^{-2}$ s$^{-1}$.
This flux (approximately 6~mCrab) is also consistent with the upper limits 
obtained by Super--AGILE in the same time periods.

%
       %%%%%%%%%%%%%%%%%%%%%%%%%%%%%%%%%%%%%%%%%%%%%%%%%%%%%%%%%%%%%%%%%%%%
                \section{Optical-to-radio data} \label{3c454:optical}
       %%%%%%%%%%%%%%%%%%%%%%%%%%%%%%%%%%%%%%%%%%%%%%%%%%%%%%%%%%%%%%%%%%%%
%
%%%%%--------------------------------------------
  \subsection{GASP-WEBT Data Reduction and Analysis} \label{3c454:webt:analysis}
%%%%%--------------------------------------------
%
The Whole Earth Blazar Telescope
(\webt)\footnote{\texttt{http://www.oato.inaf.it/blazar/webt}, see e.g.
\cite{Villata2004:WEBT:BLLac}.} has been monitoring
\source{} since the exceptional
2004--2005 outburst \citep{vil06,vil07,rai07,rai08a,rai08b},
throughout the whole period of the \agile{} observation.
We refer to \cite{rai08a,rai08b} and to \cite{Villata2009:3C454:GASP:accep}
for a detailed presentation and discussion of the radio, mm, near--IR,
optical and \swi{}/UVOT data.
%
%----------------- FIGURE 7 : GASP R band  --------------------
\begin{figure}[ht]
\resizebox{\hsize}{!}{\includegraphics[angle=-90]{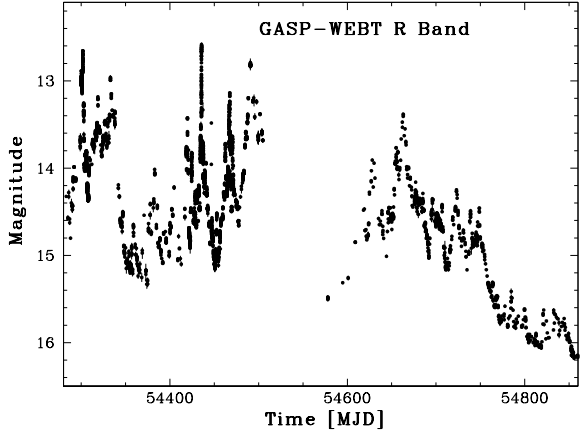}}
 \caption{GASP-WEBT light curve in the $R$ optical band 
   in the 2007--2008 and 2008--2009 observing seasons.
  }
  \label{3c454:fig:webt:lc_R}
\end{figure}
%-------------------------------------------------------------
%

\begin{figure}[ht]
\resizebox{\hsize}{!}{\includegraphics[angle=0]{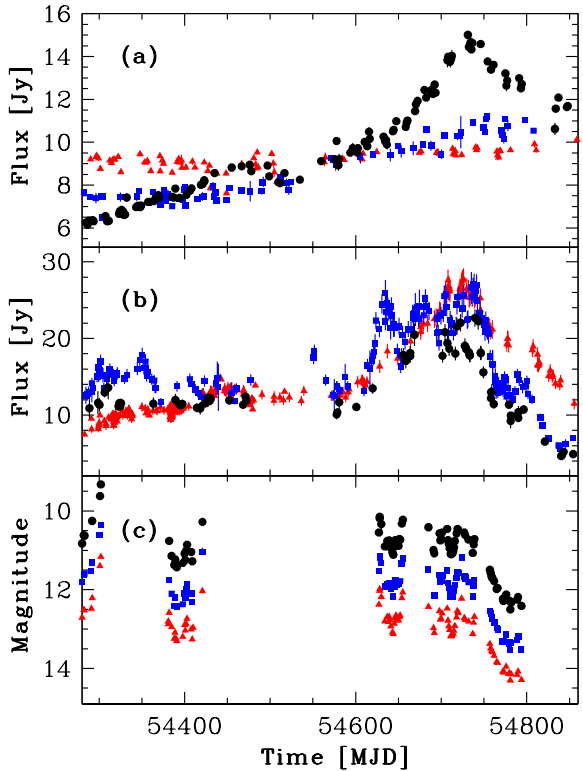}}
  \caption{{\it Panel (a)}: low-frequency radio data. Red triangles, blue
    squares and black circles represent the radio flux at 5, 8, and
    14.5 GHz, respectively.
    {\it Panel (b)}: high-frequency radio data. Red triangles, blue squares,
    and black circles represent the radio flux at 37, 230, and 345 GHz,
    respectively.
    {\it Panel (c)}: near-IR data. Red triangles, blue squares, and
    black circles, represent the $J$, $H$, and $K$ bands, respectively
    [{\it See the electronic edition of the Journal for a color version of
      this figure.}]
  }
  \label{3c454:fig:gamma_gasp}
\end{figure}

Figure~\ref{3c454:fig:webt:lc_R} shows the GASP-WEBT light curve in the
$R$ optical band, displaying several intense flares with a dynamic
range of $\sim 2.4$~mag in about 14 days,
while Figure~\ref{3c454:fig:gamma_gasp} shows
the GASP-WEBT light curves in the near-IR ($J$, $H$, $K$), radio
(5, 8, and 14.5~GHz), and mm (37, 230, and 345~GHz),
respectively\footnote{The radio-to-optical data presented in this 
paper are stored in the GASP-WEBT archive; for questions regarding 
their availability, please contact 
the WEBT President Massimo Villata ({\tt villata@oato.inaf.it}).}.
%
%
%----------------- FIGURE 8 : GASP RADIO LC  --------------------
%-------------------------------------------------------------
%

%
       %%%%%%%%%%%%%%%%%%%%%%%%%%%%%%%%%%%%%%%%%%%%%%%%%%%%%%%%%%%%%%%%%%%%
                \section{Radio VLBI data} \label{3c454:radio:vlbi}
       %%%%%%%%%%%%%%%%%%%%%%%%%%%%%%%%%%%%%%%%%%%%%%%%%%%%%%%%%%%%%%%%%%%%
%
%%%%%--------------------------------------------
  \subsection{Radio VLBI Data Reduction and Analysis} \label{3c454:radio:vlbi:analysis}
%%%%%--------------------------------------------
%

\begin{figure}[ht]
\resizebox{\hsize}{!}{\includegraphics[angle=0]{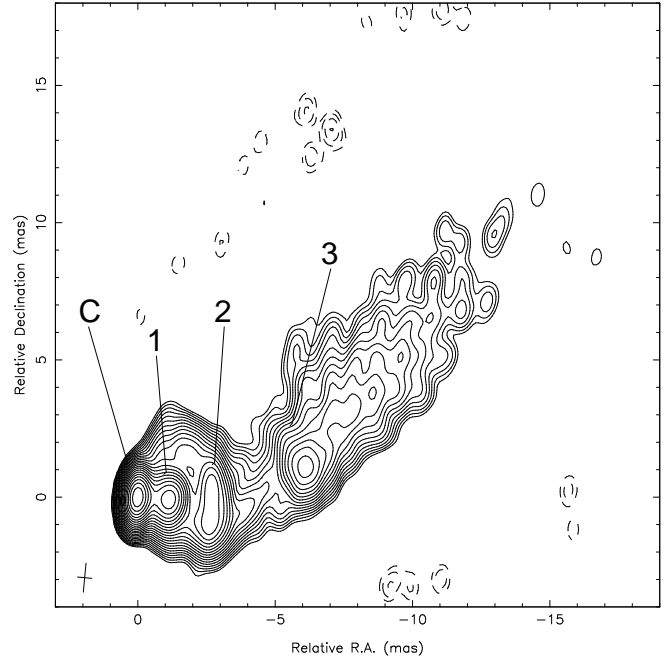}}
  \caption{VLBI image of 3C~454.3 at 15~GHz on 2007 August 9 (MJD 54321). The peak flux
    density is 2.8 Jy\,\,beam$^{-1}$ and contours are traced at $\pm (1, \sqrt(2),
    2, ...)\times1.0$ mJy beam $^{-1}$. The cross in the bottom left corner
    shows the beam FWHM, which is $1.07 \times 0.52$ mas at $-5.4$ deg.}
  \label{3c454:fig:vlbi:20070809}
\end{figure}

High resolution radio VLBI data were obtained from the MOJAVE (Monitoring Of
Jets in Active galactic nuclei with VLBA Experiments) project, a long-term
program to monitor radio brightness and polarization variations in jets
associated with active galaxies visible in the northern sky 
(\citealt{Lister2009:AJ:Mojave}; see also\footnote{\tt http://www.physics.purdue.edu/MOJAVE}).
The object was observed with the full VLBA at 15\,GHz. We obtained the
calibrated $I$ images and used the \texttt{AIPS} package to derive the position and flux
density of the core and of a few substructures in the jets using
the task \texttt{JMFIT} (Gaussian fit) (see
Figure~\ref{3c454:fig:vlbi:20070809}).
%
%----------------- FIGURE 9 : VLBI IMAGE --------------------
%-------------------------------------------------------------

%
Moreover, this source was additionally observed by VLBA at four 
epochs during the period of the maximum brightness within the 
BK150 VLBA experiment to measure parsec-scale spectra of 
\gray bright blazars (Sokolovsky et al., in preparation). We use 
15 and 43~GHz results from this program to provide better radio 
coverage of the high activity period.
These data are in agreement with MOJAVE results and
give a better statistics in the high active period.

The core is always unresolved by our Gaussian fit and uncertainties
on the flux density are dominated by calibration uncertainties (a few
percent).

\begin{figure}[ht]
\resizebox{\hsize}{!}{\includegraphics[angle=0]{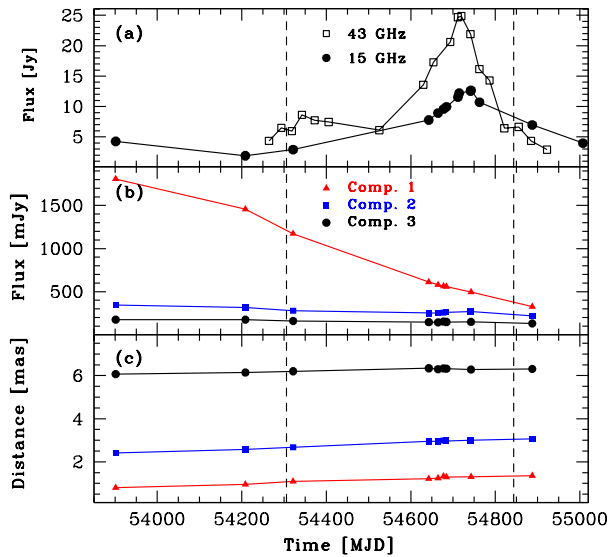}}
  \caption{Panel (a): radio core flux density at 15~GHz (filled
    circles) and at 43~GHz (open squares), respectively. Panel(b): radio components 
    flux density at 15~GHz. Panel (c): radio components motion at 15~GHz. The vertical dashed
    lines represent the start (2007 July 24) and the stop (2009
    January 12) of all \agile{} observations, respectively. 
  }
  \label{3c454:fig:vlbi:components}
\end{figure}

In Figure~\ref{3c454:fig:vlbi:components} we show the 3C~454.3 VLBI
radio core flux (panel (a))
at 15 and 43\,GHz, the radio components
flux density at 15 GHz (panel (b)), and the distance of radio components from
the core (panel (c)) as a function of time.
%
%----------------- FIGURE 10 : VLBI COMPONENTS --------------------
%-------------------------------------------------------------

The flux shows a constant increase from 2006 June 15 (MJD 53901) till 2008
October 3 (MJD 54742), followed by a fast decrease towards the last epoch
presented here, 2009 June 25 (MJD 55007).
Jet components show a well defined flux density decrease (component 1) or a
slower flux density decrease which becomes almost constant in the last epochs.
Proper motion is evident, but slowing in time for components 1 and 2; it is almost
absent for component 3.

All data are in agreement with a strong core flux density variability
possibly connected to the \gray activity, while jet components are moving
away and slowly decreasing in flux density, and are not affected by the
recent core activity. 
In  recent paper, \cite{Kovalev2009:ApJ:Gamma_Radio} indeed find a
connection between the radio and the \gray emission, correlating the
\glast{} three month data with the MOJAVE ones, and arguing that the
central region of the blazars being the source of \gray flares.
Nevertheless, a detailed study of the radio structure of \source{}
is beyond the aims of this paper, therefore the jet properties will be
discussed in depth elsewhere (Lister et al. 2009 in preparation, and Jorstad et al.
2010 in preparation).
For this reason in the following we will concentrate only on the core.

In the last two years this source has also been observed with
the VLBA at 43\,GHz (Jorstad et al. 2010; see also
\footnote{{\tt http://www.bu.edu/blazars/VLBAproject.html}}).
We used the available images to derive the core flux density of the
core at 43\,GHz. Note that, for a better comparison with 15\,GHz VLBI
data, at 43\,GHz we used natural weights and we have not searched
for possible core subcomponents (we refer to Jorstad et al. 2010 for
a detailed study of the radio structure).

The radio core shows an inverted spectrum (self-absorbed), more evident in
the high active regime, followed by a strong flux density decrease. In this
region the radio spectrum is no longer inverted.

%
       %%%%%%%%%%%%%%%%%%%%%%%%%%%%%%%%%%%%%%%%%%%%%%%%%%%%%%%%%%%%%%%%%%%%
            \section{Eighteen months of monitoring} \label{3c454:monitoring}
       %%%%%%%%%%%%%%%%%%%%%%%%%%%%%%%%%%%%%%%%%%%%%%%%%%%%%%%%%%%%%%%%%%%%
%
In this section we present a summary of all the observations on \source{}
in the period between 2007 July 24 and 2009 January 12. 
The results of the campaigns performed in
2007 July, November and December were discussed in
\cite{Vercellone2008:3C454_ApJ}, \cite{Vercellone2008:3c454:ApJ_P1}, and
\cite{Donnarumma2009:3c454:subm}, respectively.
%
%----------------- FIGURE 11 : GRID 3days 18 months  --------------------
\begin{figure}[ht]
\resizebox{\hsize}{!}{\includegraphics[angle=-90]{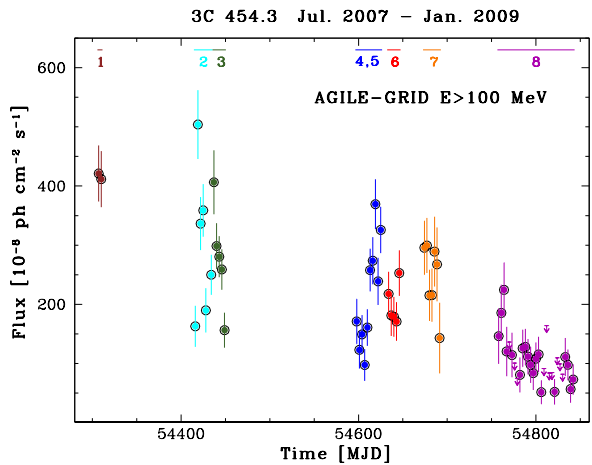}}
  \caption{\agile{}/GRID light
    curve at $\approx 3$-day resolution for E$>$100~MeV in units of
    $10^{-8}$\,\phcmsec\,. Different colors correspond to different
    observing campaigns, as described in Table~\ref{3c454:tab:grid:obs_log}. 
  }
  \label{3c454:fig:grid:3d:18months}
\end{figure}
%-------------------------------------------------------------
%

Figure~\ref{3c454:fig:grid:3d:18months} shows the \agile{}/GRID light
curve at $\approx 3$-day resolution for E$>$100~MeV in units of
$10^{-8}$\,\phcmsec. 
The light curve shows several \gray flares, with a dynamical range of
a factor of 3--4 on a time scale of about ten days. Moreover, a clear
dimming trend in the long-term light curve is present.
Table~\ref{3c454:tab:grid:flux_spec} shows the \agile{}/GRID fluxes
and spectral indices derived at different epochs.
%
%        TABLE  - GRID SPECTRAL RESULTS
%
\begin{deluxetable*}{cccc} 	
  \tablecolumns{4}
  \tabletypesize{\normalsize}
  \tablecaption{\agile{}/GRID \gray fluxes and spectral indices above 100~MeV
    at different epochs.\label{3c454:tab:grid:flux_spec}}
  \tablewidth{0pt}
  \tablehead{
    \colhead{Start Time} & \colhead{End Time} & \colhead{F$_{\rm E>100\,MeV}$}  & \colhead{$\Gamma$}\\
    \colhead{(UTC)}      & \colhead{(UTC)}    & \colhead{($\times 10^{-8}$\,ph\,cm$^{-2}$ s$^{-1}$)}  &  \colhead{} }
  \startdata
    2007-07-24 14:30   &   2007-07-30 11:40  &  $416.2 \pm  36.0$  &  $1.74 \pm 0.16$ \\
    2007-11-10 12:16   &   2007-12-01 11:38  &  $224.2 \pm  15.3$  &  $1.91 \pm 0.14$ \\
    2007-12-01 11:39   &   2007-12-16 12:09  &  $265.7 \pm  17.5$  &  $1.86 \pm 0.12$ \\
    2008-05-10 11:00   &   2008-06-09 15:20  &  $218.5 \pm  12.2$  &  $2.05 \pm 0.10$ \\
    2008-06-15 10:46   &   2008-06-30 11:14  &  $198.5 \pm  17.1$  &  $1.98 \pm 0.16$ \\
    2008-07-25 19:57   &   2008-08-14 21:08  &  $254.8 \pm  20.6$  &  $2.11 \pm 0.14$ \\
    2008-10-17 12:51   &   2009-01-12 14:30  &  $ 77.0 \pm   5.5$  &  $2.21 \pm 0.13$ \\
  \enddata
\end{deluxetable*}
%

%
%%%%%--------------------------------------------
  \subsection{Multiwavelength light curves} \label{3c454:monitoring:lc}
%%%%%--------------------------------------------
%
The \agile{}/GRID wide field of view allowed for the first
time a long-term monitoring of \source{} at energies above
100~MeV. Moreover, coordinated and almost simultaneous 
GASP-WEBT and \swi{} observations provided invaluable
information on the flux and spectral behavior from
radio to X-rays.

\begin{figure*}[ht]
\resizebox{\hsize}{!}{\includegraphics[angle=0]{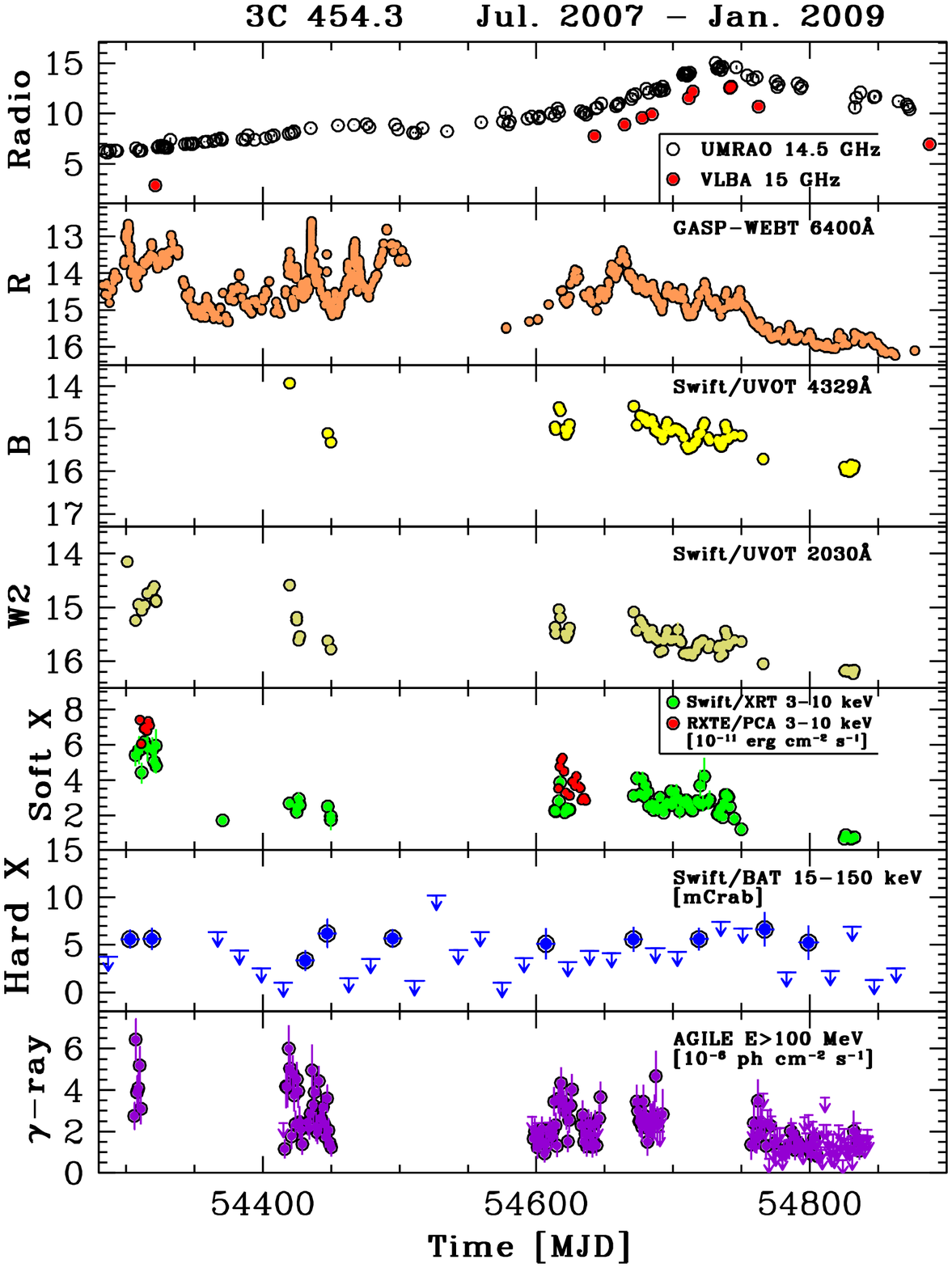}}
  \caption{\source{} light curves at different energies (see Section
    \ref{3c454:monitoring:lc} for details) over the whole time period.
  }
  \label{3c454:fig:multi:lc}
\end{figure*}

Figure~\ref{3c454:fig:multi:lc} shows the \source{} light curves 
at different energies over the whole period. 
%
%----------------- FIGURE 12 : MULTI-LC 18 months  --------------------
%-------------------------------------------------------------
%

The different panels
show, from bottom to top, the \agile{}/GRID light curve at
$\approx 1$-day resolution for E$>$100~MeV in units of
$10^{-8}$\,\phcmsec, the \swi{}/BAT light curve in the energy range
15--150~keV at $\approx 2$-week resolution, the \swi{}/XRT (filled circles)
and the \rxte{}/PCA (filled squares) light curves
in the energy range 3--10~keV , the \swi{}/UVOT light curve 
in the UV $W2$ filter, the \swi{}/UVOT light curve 
in the optical $B$ filter, the GASP-WEBT light curve 
in the optical $R$ filter, and the VLBI radio core
at 15~GHz (filled circles) and the UMRAO 14.5~GHz (open circles)
light curves, respectively.

We note that \rxte{}/PCA data are systematically higher than
\swi{}/XRT ones, which is consistent with the 20\% uncertainty
in the relative calibrations of the two instruments in this energy
band, reported by \cite{Kirsch2005:SPIE}.

\begin{figure}[ht]
\resizebox{\hsize}{!}{\includegraphics[angle=0]{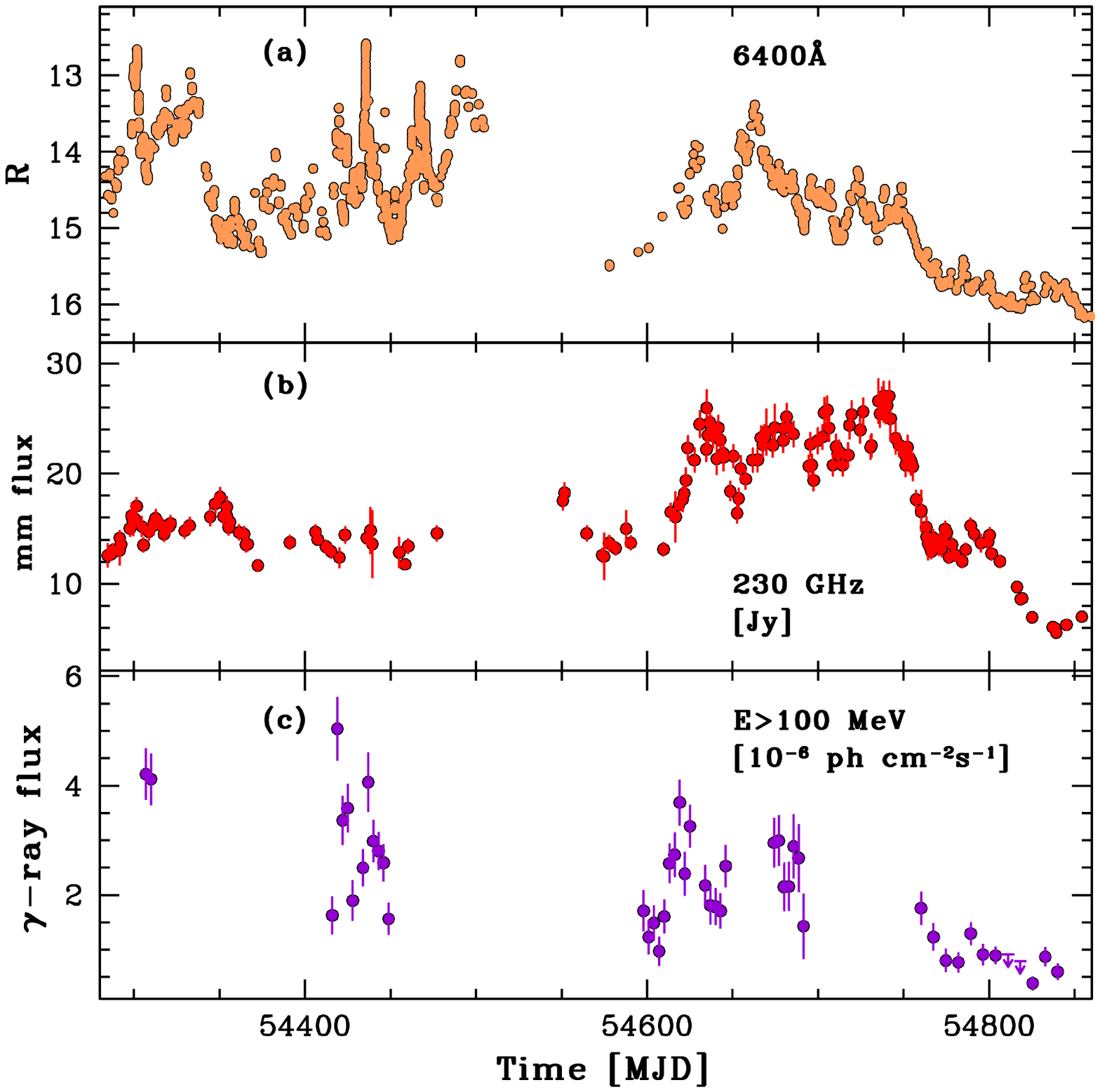}}
  \caption{Comparison between the light curves in different
    bands. Panel (a), (b), and (c) show the light curves in the
    optical, millimeter, and \gray energy bands, respectively.
  }
  \label{3c454:fig:R:mm:GRID}
\end{figure}

Figure~\ref{3c454:fig:R:mm:GRID} shows the light curves in the
$R$ band, at 1.3\,mm (230\,GHz), and above 100~MeV. The light curve in
the millimeter wavelength shows a different behavior starting
from the enhanced \gray activity at MJD$\sim 54600$, as will be
discussed in Sect.~\ref{3c454:discu:jet}.
%
%----------------- FIGURE 13 : R - 230GHz - GRID LC  --------------------
%-------------------------------------------------------------
%

%
Moreover, starting from MJD~54750, the whole jet seems to become less
energetic, with an almost monotonic flux decrease, except for a minor
burst at MJD~54800.

%
       %%%%%%%%%%%%%%%%%%%%%%%%%%%%%%%%%%%%%%%%%%%%%%%%%%%%%%%%%%%%%%%%%%%%
               \section{Discussion} \label{3c454:disc}
       %%%%%%%%%%%%%%%%%%%%%%%%%%%%%%%%%%%%%%%%%%%%%%%%%%%%%%%%%%%%%%%%%%%%

In the following Sections, we will discuss the 
correlations between the flux variations in different energy bands,
the properties of the jet, 
and the physical parameters of the emitting source. This latter point
will be addressed by means of complementary approaches, namely the SED
model fitting and discussing the geometrical properties of the jet itself.

%
%%%%%--------------------------------------------
  \subsection{Correlation analysis} \label{3c454:discu:dcf}
%%%%%--------------------------------------------
%

\begin{figure}[ht]
\resizebox{\hsize}{!}{\includegraphics[angle=0]{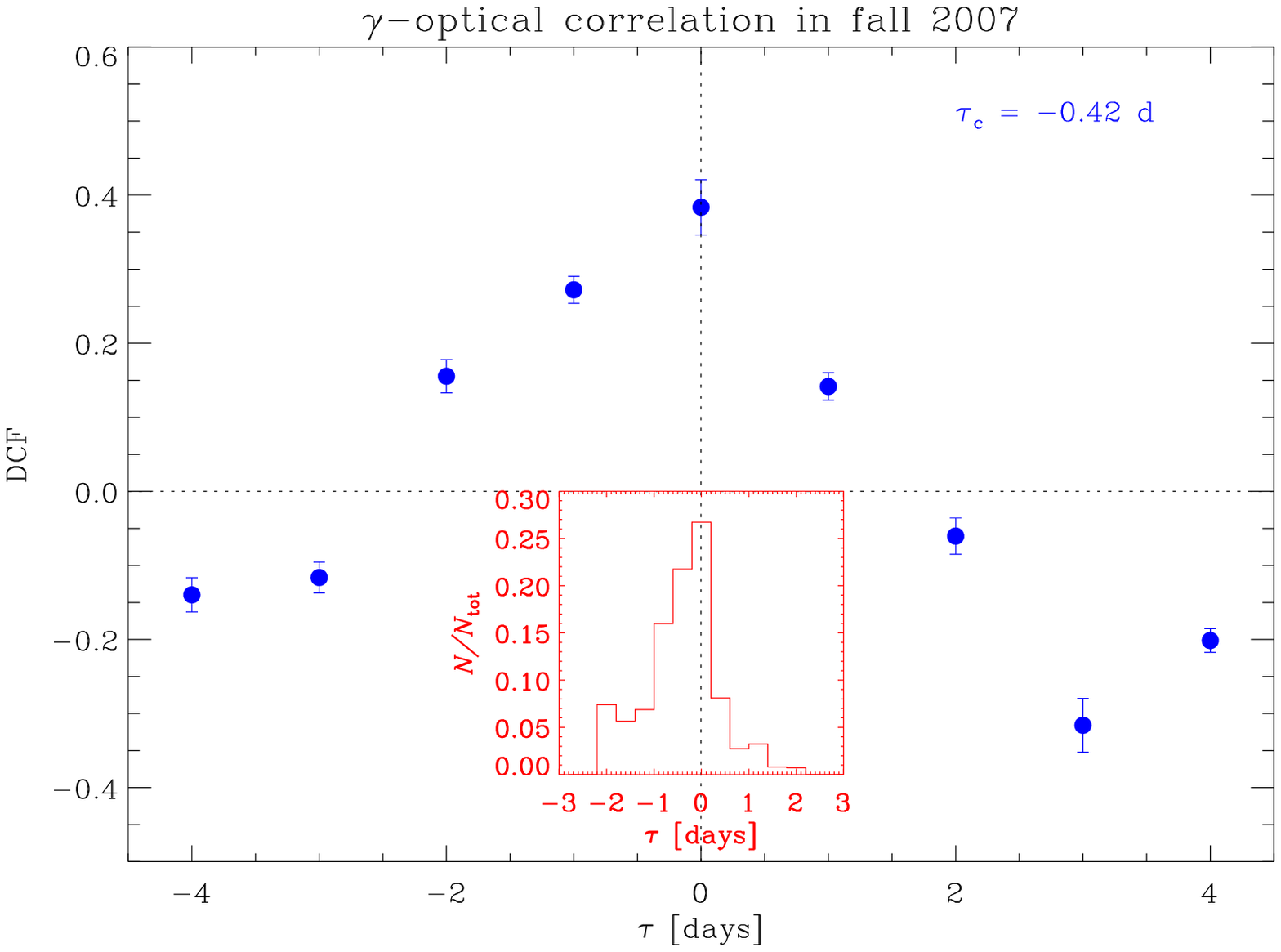}}
  \caption{Discrete correlation function between the \gray and optical
    fluxes during the ``fall 2007'' period. The uncertainty in the 
    time-lag can be computed according
    to the FR/FSS method. The inset shows the
    resulting centroid distribution (see text for details).
  }
  \label{3c454:fig:dcf_go_mc}
\end{figure}

We investigated the correlation between the $\gamma$-ray flux and the optical 
flux density in the $R$ band by means of the discrete correlation function 
\citep[DCF,][]{ede88,huf92}. This method was developed to study unevenly 
sampled data sets and can give an estimate of the accuracy of its results.
Because of the sampling gaps in the light curves, especially at $\gamma$-rays, 
we calculated the DCF on four distinct periods: July 2007 (mid 2007),
November--December 2007 (Fall 2007), May--August 2008 (mid 2008), and
October 2007 - January 2009 (Fall 2008).
The upper limits on the $\gamma$-ray fluxes were considered as detections, 
with fluxes equal to one half of the limit.
In ``mid 2007'' AGILE was pointed at 3C 454.3 when its optical main peak was 
already over; furthermore, we only have 5 $\gamma$-ray points. The low 
statistics prevents us to obtain reliable results with the DCF for this 
period.
In contrast, the period ``fall 2007'' offers a good opportunity to test the 
correlation, since the $\gamma$-ray flux, and even more the optical flux, 
exhibited strong variability. Moreover, the period of common monitoring lasted 
for more than a month. The corresponding DCF (Figure~\ref{3c454:fig:dcf_go_mc}) shows a 
maximum $\rm DCF \sim 0.38$ for a null time lag.
%
%----------------- FIGURE 14 : DCF FALL 2007  --------------------
%-------------------------------------------------------------
%

However, the shape of the peak is asymmetric, and if we calculate the 
centroid \citep{pet98}, we find that the time lag is $-0.42$\,days, 
i.e.\ about 10 hours. This result is in agreement with what was found by 
\citet{Donnarumma2009:3c454:subm} 
when analyzing the December 2007 observations only, and implies 
that the $\gamma$-ray flux variations are delayed by few hours with respect 
to the optical ones.
In the period ``mid 2008'' the main optical peak (and also the minor one) 
occurred when AGILE was not observing the source.
Finally, we computed the DCF corresponding to the ``fall 2008'' period.
We obtain a broad maximum, indicating a fair correlation
($\rm DCF_{max} \sim 0.66$), but with large errors, peaking at $-2 \, \rm day$ 
time lag, but with centroid around 0 day. This result is consistent with that 
obtained in the ``fall 2007'' period, which appears to be the most robust one.
Hence, for this case we estimated the uncertainty on the time lag by means of 
the statistical method known as ``flux randomization / random subset selection''
\citep[FR/RSS][]{pet98,rai03}. We run 2000 FR/RSS realizations and for each 
of them calculated the centroid corresponding to the maximum. The resulting 
centroid distribution shown in Figure~\ref{3c454:fig:dcf_go_mc} allows us to conclude 
that the $\gamma$-optical correlation occurs with a time 
lag of $\tau=-0.4^{+0.6}_{-0.8}$, the uncertainty corresponding 
to a $1 \sigma$ error for a normal distribution.
This result is consistent with a recent
analysis of the public \glast{} data and the optical SMARTS data 
by \cite{Bonning2008:3C454_apj}.

%
%%%%%--------------------------------------------
  \subsection{Variability analysis} \label{3c454:discu:var}
%%%%%--------------------------------------------
%
The {\it observed} variance of a light curve for a specific
detector can be written as 
\begin{equation}
S^{2} = \frac{1}{N-1}\sum_{i=1}^{N} (x_{i} - \bar{x})^{2},
\end{equation}
where $\bar{x}$ is the average value of the $x_{i}$ measurements.
Moreover, since we deal with different detectors, in order to
take into account the different count rates in different
energy bands, and to compare their variance, we consider
the normalized variance, $S^{2}/\bar{x}^{2}$.
In order to compute the {\it intrinsic} variance of a source
light curve, the measurement errors must be taken into
account, since they contribute an additional term to the 
variance.
This approach was treated in detail by \cite{Nandra1997:excvar} 
and by \cite{Edelson2002:excvar}, who introduced the term
of ``excess variance'':
\begin{equation}
\sigma_{\rm XS}^{2} = S^{2} - \bar{\sigma^{2}},
\end{equation}
where $\bar{\sigma^{2}}$ is the mean squared error,
\begin{equation}
\bar{\sigma^{2}} = \frac{1}{N} \sum_{i=1}^{N} \sigma_{i}^{2},
\end{equation}
and $\sigma_{i}$ are the measurement uncertainties of a light
curve points $x_{i}$.

Thus, the normalized excess variance, 
\begin{equation}
\sigma_{\rm NXS}^{2} = \frac{\sigma_{\rm XS}^{2}}{\bar{x^{2}}},
\end{equation}
can be used to compare variances between different observations.

In order to quantify the flux variability in different
energy bands, we computed the fractional root mean square (rms)
variability amplitude, $F_{\rm var}$, defined as
\begin{equation}
F_{\rm var} = \sqrt{\frac{S^{2} - \bar{\sigma^{2}}}{\bar{x}^{2}} }
\end{equation}
\citep[see also][and references therein]{Vaughan2003:FVAR}.
%

%
%----------------- FIGURE 15 : GRID-FVAR  --------------------
\begin{figure}[ht]
\resizebox{\hsize}{!}{\includegraphics[angle=0]{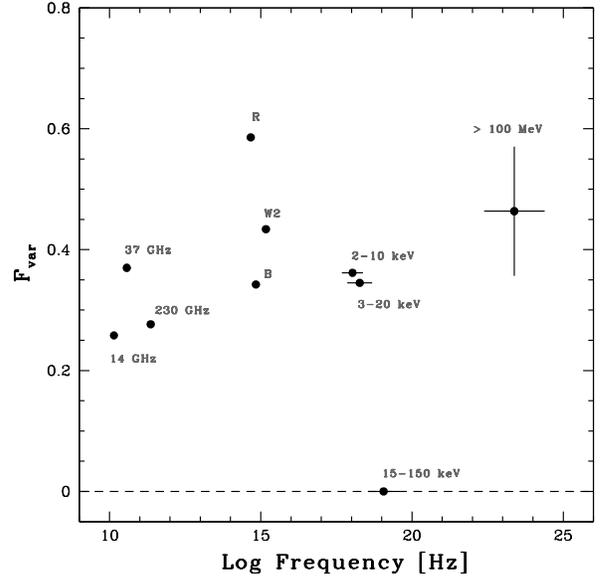}}
  \caption{Fractional rms variability amplitude as a function of
    frequency.
  }
  \label{3c454:fig:fvar}
\end{figure}
%-------------------------------------------------------------
%
Figure~\ref{3c454:fig:fvar} shows the fractional rms variability amplitude
at different frequencies.
The optical $R$ band is the one showing the highest degree of
variability. This is partly due to the higher sampling of
the optical data with respect to the the other frequencies.
Nevertheless, a possible trend (higher fractional variability
amplitude at higher frequency) is also present.
This possible trend of the fractional rms variability amplitude
with the logarithm of the frequency was observed in other sources
too \citep[see e.g. PKS~2155$-$304,][]{Zhang2005:PKS2155:fvar},
and it was interpreted as signature of spectral variability.
A more systematic study of the variability properties of
\source{} will be addressed in a forthcoming paper.

%
%%%%%--------------------------------------------
  \subsection{Radio VLBI vs \gray Data} \label{3c454:monitoring:vlbi_gamma}
%%%%%--------------------------------------------
%

Figure~\ref{3c454:fig:multi:lc} clearly shows a strong enhancement of
the radio core flux starting about on MJD~54500. The
highest flux density is on MJD~54742 at 15~GHz and on MJD~54719 at~43 GHz.
This variability is not well correlated with the variability at
higher frequencies: optical and \gray data data show more different 
flares in the period MJD 54400--54800 
(see Figure~\ref{3c454:fig:R:mm:GRID} panels (a) and (c), respectively). 
Moreover, the radio flux density increase is smooth and longer in time,
while \gray and optical flares are evolving faster.

At 230~GHz the flux density variability mimics the VLBI radio core
properties to MJD~54600, when a large flux density increase is visible,
with a peak at about MJD~54630. 
At this frequency the source remains 
in an active phase up to MJD~54700 (Figure~\ref{3c454:fig:R:mm:GRID}, panel (b)). 

This poses an interesting question as to the nature of such an increase  
of the core radio flux.
As reported in \cite{Ghisellini2007:3C454:SED} it is likely that the emitting region is more 
compact and has a smaller bulk
Lorentz factor closer to the supermassive black hole. 
We can assume that in the region active at 43~GHz, in the quiescent
state, the jet Lorentz factor is $ \Gamma \sim 10$ \citep{Giovannini2001:ApJ:VLBI}. 
To obtain the flux density
increase of the core at 43~GHz (from $\sim$ 5 Jy up to 25 Jy)
the Doppler factor has to increase up to $\delta \sim 30$.
Such an increase requires that the source is oriented at a small angle
$\theta$ with
respect to the line of sight, since 
a large change in the jet velocity will produce a small increase in the Doppler factor.
A Doppler factor $\delta = 30$ can be obtained if $\theta = 1.5^{\circ}$ and
$\Gamma = 20$, corresponding to a bulk velocity increase from 0.9950 to 0.9987
(note that a larger orientation angle, e.g. $\theta = 3^{\circ}$ with the same
increase in the jet velocity, will produce a small change in the
Doppler factor $\delta$, from 16 to 19).

The presence of one or more new jet components is not 
revealed in the high resolution VLBA images, even if the most recent VLBA 
images at 43~GHz suggest a jet expansion near to the 
radio core starting from MJD $\sim$ 54600.
Because of different properties (multiple bursts at high frequency, a single
peak in the radio band) it is not possible to correlate the radio peak
with a single \gray or optical burst. We can speculate that a multiple source
activity in the optical and \gray bands is integrated in the radio emitting
region in a single event. This event (see
Figure~\ref{3c454:fig:vlbi:components}) has a clear flux
density peak on MJD $\sim$ 54720 and we can assume that 43~GHz is the
self-absorption frequency at that epoch.
This scenario is in agreement with the one discussed by
\cite{Krichbaum2008:ASPC} (see their Figure 3).

According to \cite{Marscher1983:ApJ:SSA}, the self-absorption frequency,
the source size, flux density, and the magnetic field are correlated as follows,
\begin{equation}
B = 3.2 \times 10^{-5} \times \theta^{4}\, \nu_{\rm m}^{5}\, S_{\rm m}^{-2} \, 
\delta\, (1+z)^{-1} \,\, {\rm Gauss},
\end{equation}
where $B$ is the magnetic field in Gauss, $\theta$ the angular size in mas
(note that $\theta = 1.8 \times {\rm HPBW}$, where HPBW is the half
power beam width), $\nu_{\rm m}$ is the frequency (in GHz) of the 
maximum flux density $S_{\rm m}$ (in Jy), and $\delta$ is the Doppler
factor, respectively.
Moreover, we assume a particular value ($\alpha=0.5$) 
of spectral index in the optically thin part of the synchrotron spectrum. 

Thus, we can use the radio VLBI data at 43~GHz to constrain the physical
properties in the region where the source will start to be visible at 
at this frequency. The angular resolution in the jet direction of VLBA data at
43~GHz is $\sim$ 0.14 mas corresponding (as discussed in \citealt{Marscher1983:ApJ:SSA})
to $\theta \la 0.25$ mas.
Assuming $\delta = 30$, we obtain $B \la 0.5$ Gauss. 

It is reasonable to assume that when the source is even smaller, the emission in the
radio band is not visible being self-absorbed, and that the local
magnetic field is $ B \la 0.5$ Gauss when we start to detect the radio emission.
The size of this region should be smaller than 0.25 mas (about 2~pc).

%
%%%%%--------------------------------------------
  \subsection{Spectral analysis} \label{3c454:discu:sed}
\begin{figure}[ht]
\resizebox{\hsize}{!}{\includegraphics[angle=0]{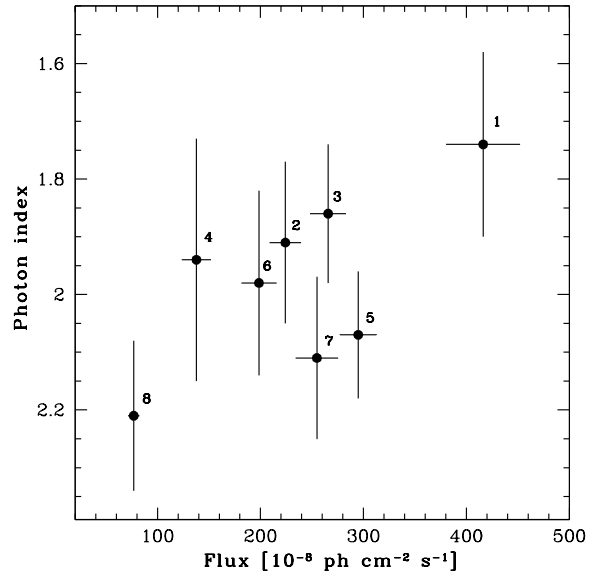}}
  \caption{\agile{}/GRID photon index as a function of the \gray flux
    above 100~MeV. Number beside each points represents the epochs
    listed in Table~\ref{3c454:tab:grid:obs_log}.
  }
  \label{3c454:fig:1yr_grid_hyster}
\end{figure}

%%%%%--------------------------------------------
%
%
The correlation between the flux level and the spectral slope
in the \gray energy band was extensively studied by means of the
analysis of the \egret{} data.
\cite{Nandikotkur2007:EGRET:slopes} showed that the behavior of EGRET
blazars is inhomogeneous.
Figure~\ref{3c454:fig:1yr_grid_hyster} shows the \agile{}/GRID 
photon index as a function of the \gray flux at different epochs.
A ``harder-when-brighter'' trend seems to be present in the long time
scale \agile{} data.
%
%----------------- FIGURE 16  : GRID  --------------------
%-------------------------------------------------------------
%

%
Further long term observations of \source{} and of other bright \gray
blazars at different flux levels with \agile{} and \glast{} will be
crucial to assess this topic.

Different emission mechanisms can be invoked to explain the \gray
emission. In the leptonic
scenario, the low--frequency peak in the blazar SED is interpreted as synchrotron
radiation from high--energy electrons in the relativistic jet,
while the high--energy peak can be produced by IC on different
kinds of seed photons. In the synchrotron
self--Compton [SSC] model (\citealt{Ghisellini1985:SSC},
\citealt{Bloom1996:SSC}) the seed photons come from the
jet itself. Alternatively, the seed photons can be those of the
accretion disk [external Compton scattering of direct disk
radiation, ECD, \cite{Dermer1992:ECD}]  or those of the broad--line
region (BLR) clouds [external Compton scattering from clouds, ECC,
\cite{Sikora1994:ECC}]. The target seed photons can also be those
produced by the dust torus surrounding the nucleus
[external Compton scattering from IR-emitting dust, ERC(IR),
\cite{Sikora2002:ERCIR}].

We fit the SEDs for the different observing periods
by means of a one-zone leptonic model, considering the contributions 
from SSC and from external seed photons originating both from the
accretion disk and from the BLR (detailed description of this model
is given in \citealt{Vittorini2009:ApJ:Model}).
Indeed, emission from both of them were detected during faint states
of the source \citep{rai07}.
%
%
%----------------- FIGURE 17 : SED MJ08  --------------------
\begin{figure}[ht]
\resizebox{\hsize}{!}{\includegraphics[angle=0]{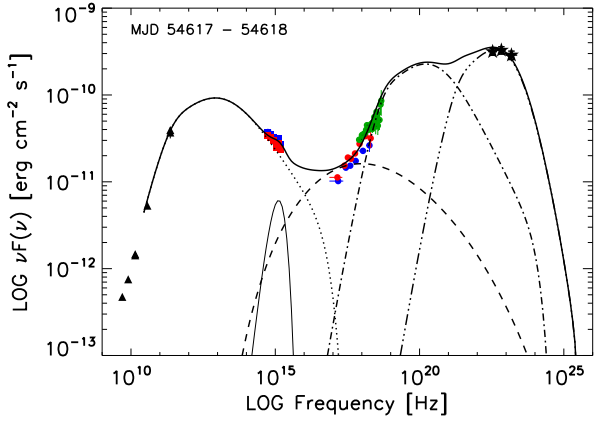}}
  \caption{\source{} SED centered on MJD~54617--54618. Black triangles, 
    red (blue) squares, red (blue) circles, green circles, and black stars represent radio,
  MJD~54617 (54618) \swi{}/UVOT, MJD~54617 (54618) \swi{}/XRT, \rxte{}/PCA, and \agile{}/GRID
  data, respectively. UV and X-ray data are de-reddened and corrected for Galactic extintion.
  The thin solid, dotted, dashed, dot-dashed, and the triple-dot-dashed,
  represent the accretion disk blackbody, the
  synchrotron, the SSC, the external Compton on the disk, and 
  the external Compton on the BLR radiation, respectively. The thick
  solid line represent the sum of all the individual components.
  }
  \label{3c454:fig:sed:mj08}
\end{figure}
%-------------------------------------------------------------
%

The emission along the jet is assumed to be produced in
a spherical blob with comoving radius $R$ by
accelerated electrons characterized by a comoving broken power
law energy density distribution of the form,
\begin{equation}
n_{e}(\gamma)=\frac{K\gamma_{\rm b}^{-1}} 
{(\gamma/\gamma_{\rm b})^{\alpha_{\rm l}}+
(\gamma /\gamma_{\rm b})^{{\alpha}_{\rm h}}}\,,
\label{eq:ne_gamma}
\end{equation}
where $\gamma$ is the electron Lorentz factor assumed to vary
between $10<\gamma<10^{4}$, $\alpha_{\rm l}$ and
$\alpha_{\rm h}$ are the pre-- and post--break electron distribution
spectral indices, respectively, and $\gamma_{\rm b}$ is the
break energy Lorentz factor. We assume that the blob
contains an homogeneous and random magnetic field
$B$ and that it
moves with a bulk Lorentz Factor
$\Gamma$ at an angle
$\Theta_{0}$  with respect to the line of sight. The
relativistic Doppler factor is then $\delta = [ \Gamma \,(1 - \beta
\, \cos{\Theta_{0}})]^{-1}$, where $\beta$ is the usual blob bulk
speed in units of the speed of light.
Our modeling of the \source{} high-energy emission is based
on an inverse Compton  model with two main sources of external
target photons:
{(1)} an accretion disk characterized by a blackbody spectrum 
peaking in the UV with a bolometric luminosity  $L_{\rm d}$
for an IC-scattering blob at a distance $r_{\rm d}=4.6 \times 10^{16}$\,cm
from the central part of the disk;
{(2)} a Broad Line Region with a spectrum peaking in the $V$ band,
placed at a distance from the blob of $r_{\rm BLR}=4 \times 10^{18}$\,cm, 
and assumed to reprocess $10\%$ of the irradiating
continuum (\citealt{Tavecchio2008:blr:cloudy,rai07,rai08b}).
%
%----------------- FIGURE 18 : SED JA08  --------------------
\begin{figure}[ht]
\resizebox{\hsize}{!}{\includegraphics[angle=0]{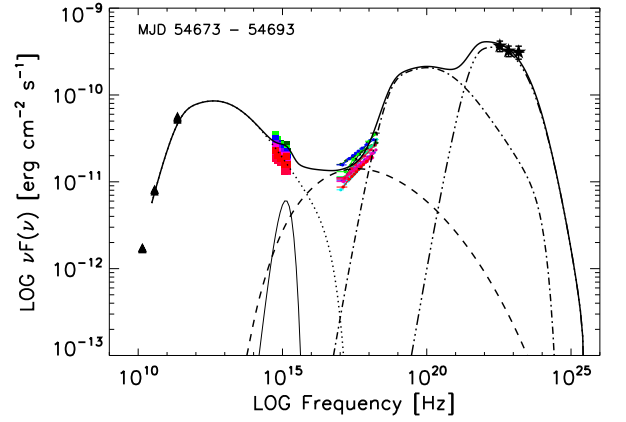}}
  \caption{\source{} SED during the period MJD~54673--54693.
    Black triangles, multicolor squares,  circles, and black stars
    represent radio, \swi{}/UVOT, \swi{}/XRT, and \agile{}/GRID
    data, respectively. UV and X-ray data are de-reddened and corrected for Galactic extintion.
  The thin solid, dotted, dashed, dot-dashed, and the triple-dot-dashed,
  represent the accretion disk blackbody, the
  synchrotron, the SSC, the external Compton on the disk, and 
  the external Compton on the BLR radiation, respectively. The thick
  solid line represent the sum of all the individual components.
}
  \label{3c454:fig:sed:ja08}
\end{figure}
%-------------------------------------------------------------
%

%
These two regions contribute to the ECD and the ECC,
respectively, and it is interesting to test the relative
importance of the two components that can be emitted by the
relativistic jet of \source{} under different conditions. We summarize here 
the main results of our best model for the different time periods.

\begin{figure}[ht]
\resizebox{\hsize}{!}{\includegraphics[angle=0]{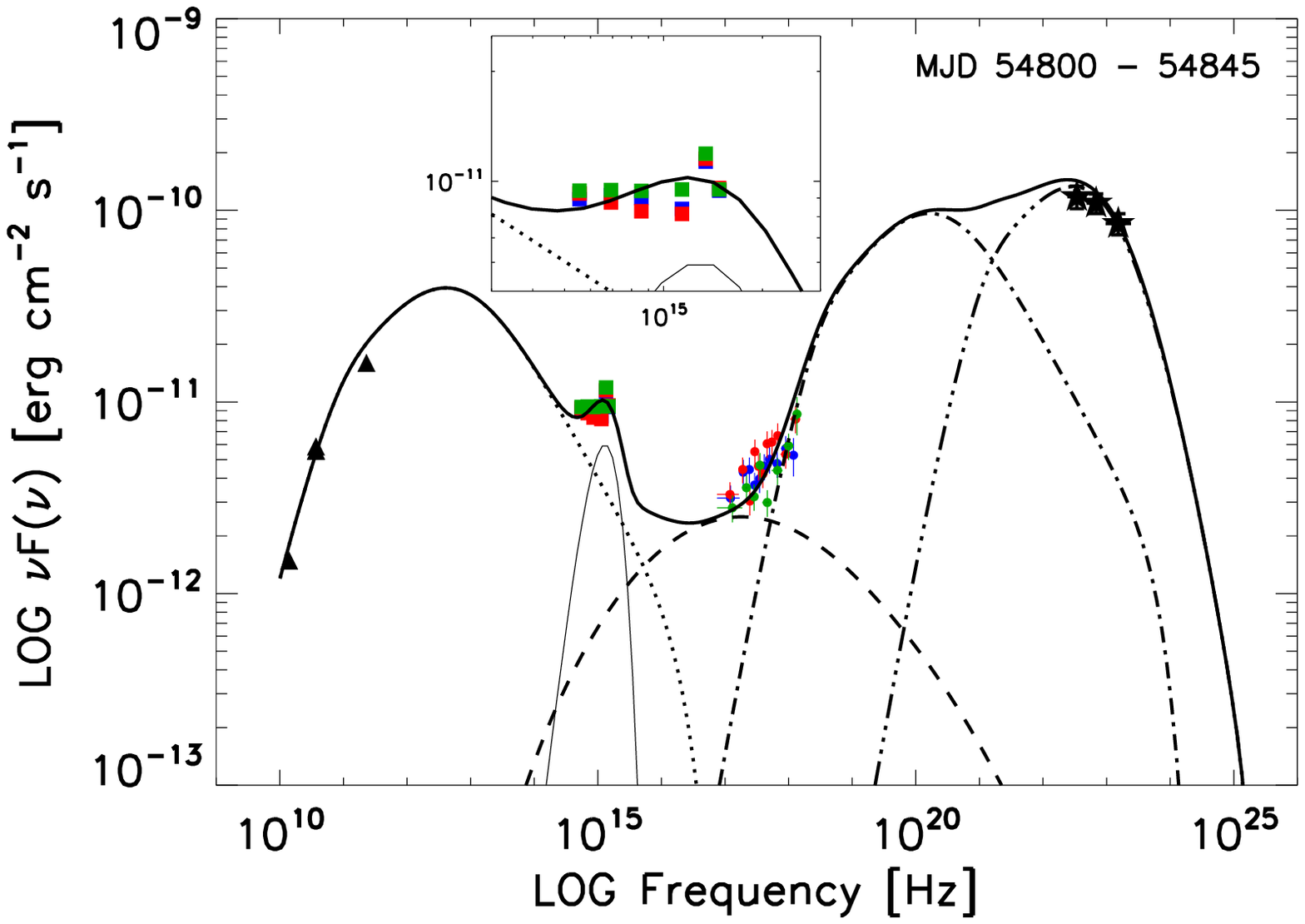}}
  \caption{\source{} SED during the period MJD~54800--54845. Black trinagles,
    multicolor squares,  circles, and black stars represent radio,
    \swi{}/UVOT, \swi{}/XRT, and \agile{}/GRID
    data, respectively. UV and X-ray data are de-reddened and corrected for Galactic extintion.
  The thin solid, dotted, dashed, dot-dashed, and the triple-dot-dashed,
  represent the accretion disk blackbody, the
  synchrotron, the SSC, the external Compton on the disk, and 
  the external Compton on the BLR radiation, respectively. The thick
  solid line represent the sum of all the individual components.
  The insert shows the portion of the SED dominated by the
  contribution of the disk blackbody radiation.
}
  \label{3c454:fig:sed:od08}
\end{figure}

Table~\ref{3c454:tab:sed:param} shows the best-fit parameters
of our modeling of SEDs corresponding to the following periods
(see Figure~\ref{3c454:figure1}): 
(SED1), MJD~54617--54618, when \source{} entered a phase of
high \gray activity;
(SED2) MJD~54673--54693, when the \gray flux 
was almost constant;
(SED3) MJD~54800--54845, when the source flux 
was at the minimum level (about $70 \times 10^{-8}$\,\phcmsec\,).
In Figures~\ref{3c454:fig:sed:mj08}, \ref{3c454:fig:sed:ja08}, 
and \ref{3c454:fig:sed:od08},
the thin solid, dotted, dashed, dot-dashed, and the triple-dot-dashed,
represent the accretion disk blackbody, the
synchrotron, the SSC, the external Compton on the disk, and 
the external Compton on the BLR radiation, respectively, while the thick
solid line represents the sum of all the individual components.
The insert of Figure~\ref{3c454:fig:sed:od08} shows the
portion of the SED dominated by the contribution of the disk blackbody
radiation, which clearly emerges since the source is a relative low state.
%
%----------------- FIGURE 19 : SED OD08  --------------------
%-------------------------------------------------------------
%

%
We find that the three SEDs can be reproduced well by very similar 
parameters, the main difference being the shape of the electron 
distribution and the break energy Lorentz factor.
We note that the observed minimum variability time scale, of the
order of half a day, is consistent with the
minimum variability time scale ($\sim 10$
hours) allowed by the model fit.
Finally, we computed for the three different SEDs the total power 
carried in the jet, $P_{\rm jet}$, defined as
\begin{equation}
P_{\rm jet} = L_{\rm B} + L_{\rm p} + L_{\rm e} + L_{\rm rad}\,\,\,{\rm
  erg}\,{\rm s}^{-1},
\label{eq:Pjet}
\end{equation}
where $L_{\rm B}$, $L_{\rm p}$, $L_{\rm e}$, and $L_{\rm rad}$ are
the power carried by the magnetic field, the cold protons, the
relativistic electrons, and the produced radiation, respectively.
We obtain a value of $P_{\rm jet}$ of $3.2 \times 10^{46}$\,erg\,s$^{-1}$, 
$3.7 \times 10^{46}$\,erg\,s$^{-1}$, and $2.5 \times 10^{46}$\,erg\,s$^{-1}$
for SED1, SED2 and SED3, respectively. 
The total power of the jet is
lower at the end of the \agile{} observing period, following the general
trend of the multiwavelength light curves.
%
%         TABLE  - SED SPECTRAL RESULT
%
\begin{deluxetable}{lrrrr} 	
  \tablecolumns{5}
  \tabletypesize{\normalsize}
  \tablecaption{Input parameters for the model of SED1, SED2, and SED3.
    See Section~\ref{3c454:discu:jet} for details.  \label{3c454:tab:sed:param}} 	
  \tablewidth{0pt}
  \tablehead{
    \colhead{Parameter} & \colhead{SED1} & \colhead{SED2} & \colhead{SED3}
    & \colhead{Units}}
  \startdata
  $\alpha_{\rm l}$             & 2.3   & 2.5   & 2.0   & \\
  $\alpha_{\rm h}$             & 4.0   & 4.0   & 4.2   & \\
  $\gamma_{\rm min}$           & 30    & 30    & 18    & \\
  $\gamma_{\rm b}$             & 300   & 280   & 180   & \\
  $K$                         & 80    & 80    & 100   & cm$^{-3}$ \\
  $R$                         & 21.5  & 21.5  & 21.5  & 10$^{15}$\,cm\\
  $B$                         & 2     & 2     & 2     & G\\
  $\delta$                    & 34    & 34    & 34    & \\
  $L_{\rm d}$                  & 5     & 5     & 5     & 10$^{46}$\,erg\,s$^{-1}$\\
  $r_{\rm d}$                  & 0.015 & 0.015 & 0.015 & pc\\
  $\Theta_{0}$                & 1.15   & 1.15   & 1.15   & degrees\\
  $\Gamma$                    & 20    & 20    & 20    & \\
  $P_{\rm jet}$                & 3.2   & 3.7    & 2.5  & 10$^{46}$\,erg\,s$^{-1}$\\
  \enddata
\end{deluxetable}

%
%%%%%--------------------------------------------
  \subsection{Jet geometry} \label{3c454:discu:jet}
%%%%%--------------------------------------------
%
The light curves in Figure~\ref{3c454:fig:R:mm:GRID} show
a different behavior starting from the end of 2007 among the
different energy bands.

As shown in \cite{Villata2009:3C454:GASP:accep}, a possible interpretation
arises in the framework of a change in orientation of a curved jet,
yielding different alignment configurations within the jet itself.

During 2007, the more pronounced fluxes and variability of the optical
and \gray bands seem to favor the inner portion of the jet as the
more beamed one. On the other hand, the dimming trend in the optical
and in the \gray bands, the higher mm flux emission and its enhanced
variability during 2008, seem to indicate that the more extended
region of the jet became more aligned with respect to the observer
line of sight.

%

       %%%%%%%%%%%%%%%%%%%%%%%%%%%%%%%%%%%%%%%%%%%%%%%%%%%%%%%%%%%%%%%%%%%%
               \section{Conclusions} \label{3c454:conc}
       %%%%%%%%%%%%%%%%%%%%%%%%%%%%%%%%%%%%%%%%%%%%%%%%%%%%%%%%%%%%%%%%%%%%
%
The \agile{} high-energy long-term monitoring of the blazar \source{} allowed us
to organize a few multiwavelength campaigns, as well as monitoring
programs at lower frequencies, over a time period of about eighteen
months. Thus, we were able to investigate the spectral energy
distributions over several decades in energy, to study the interplay
between the \gray and the optical fluxes, and the physical properties
of the jet producing the non-thermal radiation.

The {\it global} view we obtained after one and a half years of
observations can be summarized as follows:
\begin{enumerate}
\item The \gray emission for energy $E>100$\,MeV is clearly highly
  variable, on time scales of the order of one day or even shorter,
  with prominent flares reaching, on a day time scale, the order of
  magnitude of the Vela pulsar emission, the brightest, persistent 
  $\gamma$-ray source above 100~MeV.
\item Starting from October 2008, \source{} entered a prolonged
  mid- to low-level \gray phase, lasting several months.  
\item Emission in the optical range appears to be weakly correlated with that 
  at \gray energies above 100 MeV, with a lag (if present) of the \gray
  flux with respect to the optical one less than one day.
\item While at almost all frequencies the flux shows a diminishing trend with time,
  the 15~GHz radio core flux increases, although no
  new jet component seems to be detected.
\item The average \gray spectrum during the different observing
  campaigns seems to show an harder-when-brighter trend.
\item Our results support the idea the dominant emission mechanism 
  in \gray energy band is the inverse Compton scattering of external 
  photons from the BLR clouds scattering off the relativistic electrons 
  in the jet.
\item The different behavior of the light curves at different wavelengths could be
  interpreted by a changing of the jet geometry between 2007 and 2008.
\end{enumerate}
 
The simultaneous presence of two \gray satellites, \agile{} and
\glast{}, the extremely prompt response of wide-band satellites such as
\swi{}, and the long-term monitoring provided from the radio to the
optical by the GASP-WEBT Consortium will assure the chance to
investigate and study the physical properties of several blazars both 
at high and low emission states.
%

       %%%%%%%%%%%%%%%%%%%%%%%%%%%%%%%%%%%%%%%%%%%%%%%%%%%%%%%%%%%%%%%%%%%%
\acknowledgements
The AGILE Mission is funded by the Italian Space Agency (ASI) with
scientific and programmatic participation by the Italian Institute
of Astrophysics (INAF) and the Italian Institute of Nuclear
Physics (INFN). This investigation was carried out with partial
support under ASI contract N. {I/089/06/1}. We acknowledge financial 
support by the Italian Space Agency through contract 
ASI-INAF I/088/06/0 for the Study of High-Energy Astrophysics.
This work is partly
based on data taken and assembled by the
WEBT collaboration and stored in the WEBT archive at the Osservatorio
Astronomico di Torino - INAF\footnote{\texttt{http://www.to.astro.it/blazars/webt/}}.
We thank the \swi{} and RXTE Teams for making these observations possible,
particularly the duty scientists and science planners.
This research has made use of data from the MOJAVE database
that is maintained by the MOJAVE team.
The 70-cm meniscus observations was partially supported by
Georgian National Science Foundation grant GNSF/ST-08/4-404.
E.K. acknowledges support from the NCS grant No. 96-2811-M-008-033.
K.\ Sokolovsky was supported by the International Max Planck Research School 
(IMPRS) for Astronomy and Astrophysics at the universities of Bonn and 
Cologne.
We thank L. Fuhrmann for collaborating in the setup of the
\agile{}/\glast{} \swi{} monitoring campaign in the period
August--October 2008.
We also the Referee for his/her constructive comments. 

       %%%%%%%%%%%%%%%%%%%%%%%%%%%%%%%%%%%%%%%%%%%%%%%%%%%%%%%%%%%%%%%%%%%%

%
{\it Facilities:} \facility{AGILE}, \facility{{\it Swift}}, \facility{RXTE},
\facility{WEBT}, \facility{VLBA}, \facility{UMRAO}.

%%%%%%%%%%%%%%%%%%%%%%%%%%%%%%%%%%%%%%%%%%%%%%%%%%%%%%%%%%%%%%%%%%%%
%%%%%%%%%%%%%%%%%%%%%%%%%%%%%%%%%%%%%%%%%%%%%%%%%%%%%%%%%%%%%%%%%%%%
%%%%%%%%%%%%%%%%%%%%%%%%%%%%%%%%%%%%%%%%%%%%%%%%%%%%%%%%%%%%%%%%%%%%
\end{document}